\documentclass{aa}  
\usepackage[]{amsmath}  
\usepackage{graphicx}
\usepackage{color}
\usepackage{txfonts}
\usepackage{url}
\usepackage{natbib}         
\bibpunct{(}{)}{;}{a}{}{,}  
\newcommand{\vc}[1]{\mbox{\boldmath{$#1$}}}  

\begin{document} 

\title{
Reduced gas accretion on super-Earths and ice giants
}
\titlerunning{Low-mass gas envelopes}

\author{
  M. Lambrechts  \and 
  E. Lega 
} 
\institute{
Laboratoire Lagrange, UMR7293, Universit\'e C\^ote d'Azur, CNRS,
	Observatoire de la C\^ote d'Azur,  
	Boulevard de l'Observatoire, 06304 Nice Cedex 4, France\\
\email{michiel.lambrechts@oca.eu} 
}

\date{Received --- ; accepted ---}

\abstract{
A large fraction of giant planets have gaseous envelopes that are limited to
about 10\% of their total mass budget.
Such planets are present in the Solar System (Uranus, Neptune) and are
frequently observed in short periods around other stars (the so-called
Super-Earths).
In contrast to these observations, theoretical calculations based on the
evolution of hydrostatic envelopes argue that such low mass envelopes cannot be
maintained around cores exceeding five Earth masses.
Instead, under nominal disc conditions, these planets would acquire massive
envelopes through runaway gas accretion within the lifetime of the
protoplanetary disc.
In this work, we show that planetary envelopes are not in hydrostatic balance,
which slows down envelope growth.
A series of $3$-dimensional, global, radiative hydrodynamical simulations reveal
a steady state gas flow, which enters through the poles and exits in the disc
midplane. 
Gas is pushed through the outer envelope in about $10$ orbital timescales.
In regions of the disc that are not significantly dust-depleted, envelope
accretion onto cores of about five Earth masses can get stalled as the gas flow
enters the deep interior.
Accreted solids sublimate deep in the convective interior, but small
opacity-providing grains are trapped in the flow and do not settle, which
further prevents rapid envelope accretion.
The transition to runaway gas accretion can however be reached when cores grow
larger than typical Super-Earths, beyond $15$ Earth masses, and preferably when
disc opacities are below $\kappa=1$\,cm$^{2}$/g.
These findings offer an explanation for the typical low-mass envelopes around
the cores of Super-Earths.
}

\keywords{
  Planets and satellites: formation --
  Planets and satellites: gaseous planets --
  Hydrodynamics -- 
  Methods: numerical
  }

\maketitle

\section{Introduction}

In the core accretion scenario, giant planets form by attracting a
gaseous envelope onto a previously formed solid core \citep{Pollack_1996}. 
These large cores of a few Earth masses (M$_{\rm E}$) form from planetary
embryos that are able to accrete larger than km-sized planetesimals
\citep{Rafikov_2004} and sweep up cm-sized pebbles with high efficiency
\citep{Lambrechts_2012}.
As the core grows large, the envelope is accreted from the gaseous component
of the protoplanetary disc, which only remains present for a few million years
around young stars \citep{Ribas_2015}.
Core formation must therefore proceed at a rate of several Earth
masses per $10^6$\,year.

The accretion of the gaseous atmosphere itself can be divided in two separate
phases.
Early on, the envelope slowly gains mass as it cools down, a process which can
be further delayed by continued heat deposition from solid accretion. 
However, if the cooling process can proceed efficiently and for a sufficient
amount of time, the envelope can grow as massive as the core. 
This process triggers a second phase of rapid gas accretion.
The onset of a self-gravitating gas envelope facilitates continued
envelope cooling and mass growth. Planets that enter this regime become
gas giants with a high-mass envelope.

In this work we focus on planets in the process of attracting a gaseous
envelope, before reaching the point of runaway gas accretion.  
This growth phase has been studied extensively with 1D numerical models that
assume a spherical symmetric atmosphere and evolve the envelope assuming
hydrostatic equilibrium at all times, a method with origin in stellar evolution
calculations \citep{Mizuno_1980,Ikoma_2000,Piso_2014}. 
Such models show that the timescale to reach runaway gas accretion is very
sensitive to 
the solid accretion rate, 
the core mass, 
the dust opacity in the outer envelope, and the
amount of dissolved solids \citep{Stevenson_1982}.
Recent work has argued that the dust grains do not contribute much to the
opacity, because they coagulate and settle rapidly \citep{Ormel_2014}. 
This reduces the timescale for which a completed $10$-Earth mass core starts
runaway gas accretion to $\sim1$\,Myr, which is shorter than the disc lifetime
\citep{Mordasini_2014}.
Similarly, the addition of sublimated solids raises the mean molecular weight of
the envelope and even further reduces the time to runaway gas accretion
\citep{Hori_2011,Venturini_2015,Venturini_2016}.
Together these findings hint that all cores above a few Earth masses should have
massive gas envelopes.

These theoretical findings are however in conflict with the existence of the ice
giants in the Solar System: Uranus and Neptune.
Both planets are larger than $10$\,M$_{\rm E}$, but  never accreted a massive
gas envelope.
It also does not agree with the characterized population of short-period
exoplanets. 
Super-Earths, planets with sizes between $1$--$4$ Earth radii (R$_{\rm E}$), are
found around about half of the sun-like stars \citep{Winn_2015}.
Also these planets, with masses up to about $25$\,M$_{\rm E}$, typically have
hydrogen/helium envelopes that do not exceed $10$\% of the total mass of the
planet \citep{Hadden_2016}.

New hydrodynamical simulations challenge the standard assumptions used in
envelope growth models.
Previous evolution studies \citep{Pollack_1996,Ikoma_2000,Piso_2014}
assumed envelopes in hydrostatic balance that are closed systems. 
Interaction with the protoplanetary disc would only occur through an imposed
outer boundary condition that matches the unperturbed protoplanetary disc.
Instead, hydrodynamical simulations performed in three dimensions show that gas
continuously flows from the poles through the planetary envelope
\citep{Ayliffe_2012, Tanigawa_2012, Angelo_2013}.
Moreover, recent simulations that were performed under the simplifying
assumption of a constant temperature envelope and disc, quantified that gas is
continuously advected through the envelope on orbital timescales, implying that
almost no gas accretion would occur \citep{Ormel_2015, Fung_2015}.
However, the simplifying assumption of an isothermal envelope corresponds to
the limit case of a completely cooled-down planet, which cannot be obtained for
planets heated by the process of formation.
Therefore, these new results are difficult to interpret. 
Furthermore, hydrodynamical studies in 3D that have a more complete
thermodynamical treatment of the envelope, claim results well in line with
envelope growth rates from standard hydrostatic models \citep{Angelo_2013}.

Here, we aim to find find the quasi-static structure of a low-mass
envelope around a massive core ($5$--$15$\,M$_{\rm E}$).
We do not make an isothermal approximation, but instead use a radiative
treatment similar to \citet{Angelo_2013}. 
Additionally, we model the heat deposited by solid accretion.
In order to correctly describe disc-planet interaction, we model a full annulus
of the protoplanetary disc around the planet.
Hydrodynamical simulations covering the complete evolution of envelope
growth in time are numerically unfeasible. 
Therefore, we make temporal snapshots at different growth stages that
show how envelopes are structured and grow.

The paper is structured as follows.
A technical description of the methods can be found in Section \ref{sec:methods}.
We first study the case of a planet with a luminosity that is dominated by
the heat released by the accretion of solids (Sec.\,\ref{sec:resL27}).
This case is investigated in detail to establish the dynamical and energetic
structure of planetary envelopes in their formation stage.
Then, we proceed by studying planetary envelopes
after solid accretion has come to a halt
(Sec.\,\ref{sec:resL0}).
Our results depend on the opacity in the envelope and the mass of the solid core
of the planet.
High dust opacities tend to delay envelope growth (Sec.\,\ref{sec:depop}).
Large cores facilitate the transition to runaway gas accretion
(Sec.\,\ref{sec:depMc}).
In Section \ref{sec:Imp}, we summarize the implications of our work on the growth
of the envelope and the core. 
We place our results in the context of exoplanet observations that reveal a high
occurence rate of Super-Earths, but argue for a low frequency of more massive
gas giants
(Sec.\,\ref{sec:occ}).
Finally, a summary is presented (Sec.\,\ref{sec:Sum}).

\section{Methods}
\label{sec:methods}

\begin{figure}[t!]
  \centering
  \includegraphics[width=8.8cm]{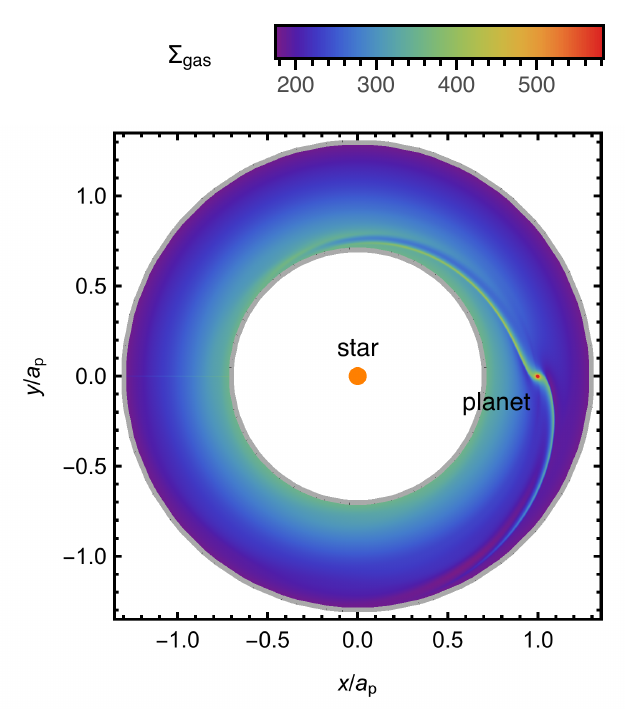}
  \caption{
  Global view on the vertically-integrated density ($\Sigma_{\rm gas}$ in
  g/cm$^2$) of the simulated annulus, 
  taken from run \texttt{RADL27k0.01}.
  We can identify the spiral arms originating from the planet located at
  coordinates $(1,0)$ revealing a strong interaction between planet and disc.
  }
  \label{fig:global}
\end{figure}

\begin{table*}
  \caption{
Main parameters for all the simulation that are listed in column 1.
The following columns give 
the gas surface density $\Sigma_{\rm p}$ 
at the position of the planet in units of $M_*/a_{\rm p}^2$, 
viscosity $\nu$ in units of $a_{\rm p}^2\Omega_{\rm p}$, 
the aspect ratio $H/r$ at the position of the planet, 
the core mass $M_{\rm c}$ in Earth mass, 
the accretion luminosity $L_{\rm acc}$ in erg/s,
the opacity $\kappa$ in cm$^2$/g and  
the smoothing length $r_{\rm smooth}$ as fraction of the Hill radius.
The second-to-last column gives the number of orbits at the position
of the planet, that were performed at the
corresponding resolution level listed in the last column (see
Sec.\,\ref{sec:NumIm} for a detailed description of the non-uniform grid that
was used).
}
\label{tab:runs}
\centering
\begin{tabular}{c c c c c c c c c c c}  
\hline\hline  
Name  & 
$\Sigma_{\rm g}$ & 
$\nu$ & 
$H/r$ & 
$M_{\rm c}$ [M$_{\rm E}$] &  
L$_{\rm acc}$ [erg/s] & 
$\kappa$ [cm$^2$/g] &  
$r_{\rm smooth}$ &
N$_{\rm orb}$ &
Resol.
\\ 
\hline

\texttt{RADL27k0.01}   & $1.25\times 10^{-3}$ & $10^{-5}$  & 0.03 & 5 &
$10^{27}$  & 0.01 & 0.125 & 30,15,4   &  low,mid,high \\

\texttt{RADL0k0.01}  & $1.25\times 10^{-3}$ & $10^{-5}$  & 0.03 & 5 & --
& 0.01  & 0.125 &   30,15,4   &  low,mid,high  \\

\texttt{RADL27k1}   & $1.3\times 10^{-3}$ & $10^{-5}$  & 0.05 & 5 &
$10^{27}$  & 1  & 0.125  &   30,8,4   & low,mid,high  \\

\texttt{RADL0k1}   & $1.3\times 10^{-3}$ & $10^{-5}$  & 0.05 & 5
& --  & 1 & 0.125 &  30,8,4    &  low,mid,high \\

\texttt{RADL27k1M15}  & $1.3\times 10^{-3}$ & $10^{-5}$  & 0.05 & 15 &
$10^{27}$  & 1 & 0.125 &  30,16,4    &  low,mid,high \\

\texttt{RADL0k1M15}   & $1.3\times 10^{-3}$ & $10^{-5}$  & 0.05 & 15 &
--  & 1 & 0.125  &   30,16,4   &  low,mid,high \\

\texttt{RADL027k0.01LONG}   & $1.3\times 10^{-3}$ & $10^{-5}$  & 0.03 & 5
& $10^{27}$  & 0.01 & 0.25 & 30,50     &  low,mid \\

\texttt{RADL0k0.01LONG}   & $1.3\times 10^{-3}$ & $10^{-5}$  & 0.03 & 5 &
--  & 0.01  & 0.25 &  30,50    &  low,mid \\

\texttt{ISO}   & $6.8\times 10^{-4}$ & $10^{-6}$  & 0.03 & 5 & --
 & -- & 0.1 &  25   &  mid \\

\hline
\end{tabular}
\end{table*}

\subsection{General description of the model}

We  model a full annulus encompassing the orbit of the planet, as illustrated in
Fig.\,\ref{fig:global}. 
This allows us to carefully analyze the exchange of gas between the envelope and
the disc.
The simulations are performed in 3D, as gas motion perpendicular to the
midplane regulates the gas dynamics around the envelope.
We use a non-uniform grid to be able to trace the flow of gas on small scales
well inside the Hill sphere of the planet, following the approach by
\citep{Fung_2015}.
Additionally we include the heat deposited by the accretion of solids, and model
energy transfer by radiation, similar to \citet{Angelo_2013}.
In order to do so, we make use of the \texttt{FARGOCA} code \citep{Lega_2014}. 
This radiation-hydrodynamics code is designed for global 3D studies of
planet-disc interaction.
It is based on the finite difference code \texttt{FARGO} \citep{Masset_2000}. 
Below, we report in more details the methods used to model the planetary
envelope.

\subsection{Gas dynamics}

We simulate a non-selfgravitating viscous gas on a fixed grid, co-rotating with the planet . 
The standard continuity and momentum equations, 
\begin{align}
  &\frac{D \rho}{Dt}   + \rho( \nabla \cdot \vc u ) = 0  \\
  &\frac{D \vc u}{Dt} = -\frac{1}{\rho} \nabla P - \nabla \Phi + \frac{1}{\rho}
  \nabla \cdot \mathbb{T} \,,
\end{align}
are solved using spherical coordinates (see \citet{Lega_2014} for more details
on the implementation in \texttt{FARGOCA}).
Here, $D/Dt$ is the Lagrangian time derivative,  $\rho$ is the density, $P$ is
the gas pressure, $\vc u$ is the velocity and $\mathbb{T}$ is the viscous
stress tensor. 

The gravitational potential $\Phi$ is the sum of the contribution of the stellar
potential, the potential of the planet and indirect terms that take into account
the acceleration of the primary due to the gravity of the planet and the disc.
The planetary potential is given by $\Phi_{\rm p} = -GM_{\rm c}/r$ outside a
radius $r_{\rm sm}$ form the planet. Here $G$ is the gravitational
constant and $M_{\rm c}$ is the core mass. 
Within the smoothing length $r_{\rm sm}$, the central singularity is avoided
by a cubic interpolation to $\Phi_{\rm p} =-2GM_{\rm c}/r_{\rm sm}$
\citep{Klahr_2006}.
In appendix \ref{ap:res}, we describe this prescription for the planetary
potential in more detail and explore its influence on our results.

\subsection{Gas thermodynamics}

\subsubsection{Energy equation}
\label{sec:Eeq}

We separately evolve the thermal energy density ($ E_{\rm gas}$) and radiative
energy density ($ E_{\rm rad}$).
The evolution equations are given by \citep{Kley_1989, Commercon_2011,
Bitsch_2013}:
\begin{align}
  &\frac{\partial E_{\rm rad}}{\partial t} + \nabla \cdot \vc F =
   \mathcal{I}(E_{\rm rad},T) \\
  &\frac{\partial E_{\rm gas}}{\partial t} 
  +  \nabla \cdot(E_{\rm gas} \vec  u) =
  -P \nabla \cdot \vc u - \mathcal{I}(E_{\rm rad},T) + Q^+  +l\,.
  \label{eq:Edot}
\end{align}
Here, $\vc F$ is the radiative flux.
The fluid-radiation interaction term is described by $\mathcal{I} = \rho \kappa
(4\sigma_{\rm SB} T^4 - cE_{\rm rad})$, where $\sigma_{\rm SB}$ is the
Stefan-Boltzmann constant, $\kappa$ is the Rosseland mean opacity, $T$ is
the temperature and $c$ is the speed of light.
In the optical thick limit, this term reduces to energy diffusion of the form
$D \nabla^2 E_{\rm gas}$. The diffusion coefficient, 
\begin{align}
  D = \frac{1}{3} c \frac{1}{\kappa \rho} 
  \left( \frac{(16\sigma_{\rm SB}T^3)/c}{c_{V} \rho}\right) \,,
  \label{eq:Ddiff}
\end{align}
corresponds to the product of the velocity and the mean free path of the
photons and the last factor in brackets weights the expression with the ratio
between the specific heat of the photons with respect to the gas ($c_{ V}$).
The term $P \nabla \cdot \vc u$ in Eq.\,\ref{eq:Edot} represents compressional
heating and $Q^+ $ viscous heating
\citep[for more detail see][]{Mihalas_1984}.
The final term, $l$, is a source term which represents the deposited heat per
unit time by solid accretion. The implementation of this term is discussed in
more details in Appendix \ref{ap:L}.

The radiative flux is calculated in the framework of the flux-limited diffusion
approximation, 
\begin{align}
  \vc F  = -\frac{c f_\lambda}{\rho \kappa}  \nabla E_{\rm rad}\,,
\end{align}
where $f_\lambda$ is a flux-limiter \citep{Levermore_1981}. 
The implementation is described in more detail in \citet[][]{Lega_2014}. 

\subsubsection{Opacity}
Both gas and dust act as a source of opacity which absorbs and deflects
radiation.
Below temperatures of around $1700$\,K dust particles are the dominant
opacity source \citep{Bell_1994}. 
In this work, we purposely choose a simple constant opacity prescription for our
envelope and gas disc.
We then explore this free parameter in Section \ref{sec:depop}.

\subsubsection{Equation of state}

Similar to our choice of a constant opacity, we have a employed a simple
equation of state. 
We use an ideal gas, where the pressure is given by
\begin{align}
  P = \frac{R}{\mu} \rho T, 
\end{align}
Here $R$ is the gas constant.
We used a fixed value for the mean molecular weight ($\mu=2.35$),
corresponding to a solar H/He mixture. 
More realistic prescriptions would need to take into account the pollution of
the envelope by heavy elements and the changes to the mean molecular weight 
and adiabatic index.
Additionally, variables like the specific heat at constant volume $c_V$ are
temperature sensitive and depend on the unknown ratio of ortho- to parahydrogen
\citep{Angelo_2013}.
These topics can be investigated in future work.

\subsection{Numerical implementation}
\label{sec:NumIm}

\subsubsection{Coordinate system}

We use spherical coordinates $(r,\theta, \varphi)$ where $r$ is the radial
distance from the origin, $\theta$ the polar angle measured from the $z$-axis
and $\varphi$ is the azimuthal coordinate starting from the $x$-axis. 
The midplane of the disc is at the equator $\theta = {\pi \over 2}$ and the
origin of the coordinates is centered on the star.
We work in a coordinate system which rotates with angular velocity, 
\begin{align}
\Omega_{\rm p} = \sqrt {G(M_*+M_{\rm c}) \over {a_{\rm p}}^3} \simeq \sqrt
{GM_* \over {a_{\rm p}}^3}\,.
\end{align}
The code operates in code units with the stellar mass $M_*$, the semi-major axis
of the planet $a_{\rm p}$ and the Keplerian frequency set to unity.

\subsubsection{Simulation domain and boundary conditions}

Radially, the disc extends from $a_{\rm min}/a_{\rm p}=0.7$ to $a_{\rm
max}/a_{\rm p}=1.3$. 
In the vertical direction the disc extends from the midplane ($\theta \simeq
90^{\circ} $ to $6^{\circ}$ above the midplane, i.e. $\theta \simeq 84^{\circ}
$). We do not study inclined planets orbits therefore we do not need to extend
the disc below the midplane. 

We use periodic boundary conditions in the azimuthal direction, as we aim
to simulate the full $2\pi$-annulus.
We use a reflective boundary on the midplane. This is appropriate for
non-inclined planets, where mirror symmetry can safely be assumed between
the upper and lower hemisphere of the disc. The upper boundary condition is
also reflective, to prevent mass outflow.
In the radial direction we use evanescent boundaries with a wave suppression
zone covering $10$\% of the inner and outer radius, to prevent the reflection of
density waves \citep{deVal_2006}. 

\subsubsection{Disc model}

Before inserting the planet, we bring the disc in radiative equilibrium
\citep{Lega_2015}.
This equilibrium is obtained for a 2D ($r,z$) axisymmetric disc with resolution 
$(N_r,N_\theta,N_\varphi)= (200,52,2)$ 
in about 100\,$\Omega_{\rm p}^{-1}$.  

Viscous stresses are the only heating source, as we ignore stellar heating.
This is a good description for the inner disc, where viscous heating dominates.
We have used a constant viscosity $\nu=10^{-5} a_{\rm p}^2
\Omega_{\rm p}$ (code units). 
This low viscosity will be necessary to properly capture the 3-dimensional flow.
We noticed higher values of viscosity (of the order $\nu=10^{-4} a_{\rm p}^2
\Omega_{\rm p}$) effectively force 2-dimensional flow, in agreement with
\citet{Fung_2015}.
We also verified that extremely low values of $\nu=10^{-6} a_{\rm p}^2
\Omega_{\rm p}$ do not alter our results significantly, but at times may
produce unsteady fluctuations. 

The gas surface density follows a power law profile, $\Sigma_{\rm gas} =
\Sigma_{\rm p} (a/a_{\rm p})^{-1/2}$\,g/cm$^2$, which gives an equilibium disc
without radial gas motion.
At the location of the planet, we record a surface density of 
$\Sigma_{\rm p} = 1.3 \times 10^{-3}$ \,$M_{*}/a_{\rm p}^2$, corresponding to
$\Sigma_{\rm p} = 400$\, g/cm$^2$ at a distance of $5.2$\,AU.
The aspect ratio is the result of the equilibrium  between viscous heating and
radiative cooling. 
Depending on the choice of the opacity, we find a gas scale aspect
ratio of $H/a_{\rm p} =0.03$ for $\kappa=0.01$\,cm$^2$/g or $H/a_{\rm p} =0.05$
for $\kappa=1$\,cm$^2$/g at the position of the planet.
We note however that our results on the small scale of planetary envelopes only
weakly depend on the exact choice of the large scale disc structure.

\subsubsection{Introduction of the planet}

After the procedure to generate an equilibrium disc, we refine the grid
azimuthally and work in three dimensions. 
Planets of $5$ or $15$ Earth masses are then embedded in the equilibrium disc
and held on fixed circular orbits in the midplane, with code units $r_{\rm
p}=1$, $\varphi_{\rm p} = 0$, $\theta_{\rm p}=\pi/2$.
Our mesh is chosen so that the planet is at the intersection between cell's
surfaces in the 3 direction, i.e. at the corner of 8 grid cells.

Since hydrodynamical 3D calculations of fully radiative discs are 
expensive in computational time, we split our simulations in 2 phases. 
In the first phase, we employ a uniform grid so that we can use the FARGO
algorithm \citep[Fast Advection in Rotating Gaseous Objects,][]{Masset_2000}
and benefit from large time steps. We recall that in a Keplerian disc the
traditional Courant condition provides very small time steps due to the fast
orbital motion at the inner boundary of the disc. In the FARGO algorithm, the
time step is instead limited by the perturbed azimuthal velocity with respect
to the Keplerian one. 
We run each simulation with the uniform grid until the disc relaxes to the
presence of the planet. This occurs after
approximately\,$30$\,$\Omega_{\rm p}^{-1}$.
Fig.\,\ref{fig:global} shows the gas surface density when the stationary state is reached.
For this uniform grid, our nominal resolution corresponds to
$(N_r,N_\theta,N_\varphi)= (200,52,2048)$ grid cells. 
This choice provides a resolution of $0.003$\,$a_{\rm p}$, corresponding to 
10 gridcells in the Hill sphere of a $5$\,M$_{\rm E}$ planet.

\subsubsection{Increased resolution around the planet}
In the second phase, we use a non-uniform grid with increasing resolution
towards the planet to more accurately model the planetary envelope.
The grid layout is based on the formulation by \citet{Fung_2015}, which gives 
small nearly uniform grid cells in the Hill sphere around the planet
and large grid cell sizes further out. 

Our medium resolution corresponds to a width of the inner grid cells
of $0.00135$\,$a_{\rm p}$, which resolves the Hill sphere of a 5\,M$_{\rm E}$
planet by about $24$ grid cells.
In total the grid has $(N_r,N_\theta,N_\varphi)= (200,52,1512)$ number of cells
in the radial, polar and azimuthal direction.
At this resolution we have about 1 orbit in 8 hours of computation with 60
cores, using hybrid MPI+OpenMP parallelization.
We typically run the code for at least $8$ orbits to relax the system to
equilibrium.

In high resolution runs, we use inner grid cells of width
$4.27\times10^{-4}$\,$a_{\rm p}$, corresponding to about $84$ grid cells across
a Hill sphere for a 5\,M$_{\rm E}$ planet.
This finer grid has $(N_r,N_\theta,N_\varphi)= (300,76,2268)$ number of cells in
the radial, polar and azimuthal direction.
At this resolution we have about $1$ orbit in $50$ hours of computation with
$60$ cores. 
In this final step, we simulate $4$ planetary orbits, which we found to be
sufficient to capture the interior structure of the planet.

Finally, we note that the FARGO advection algorithm cannot be used on the
nonuniform grid; instead the time step is set by the usual Courant condition
\citep{Stone_1992}.

\subsection{Simulations}

Our main numerical experiments are described in detail in Sections
\ref{sec:resL27}, \ref{sec:resL0}, \ref{sec:depop} and \ref{sec:depMc}.
Table \ref{tab:runs} gives an overview of the performed simulations and the
parameter space covered.
Additionally, we performed several performance tests of the code.
In Appenix\,\ref{ap:res}, we present several convergence tests. We investigate
the dependency on the resolution in the code, with a focus on properly resolving
the planetary potential.
Additionally, long-term integrations are performed to demonstrate steady-state
stability.

\section{Results: The envelope around a core accreting solids}
\label{sec:resL27}

\subsection{Overview}

We start with a presentation of our results on the envelope around a planetary
core in the process of accreting solids (run \texttt{RADL27k0.01}).
The structure of the envelope is discussed in detail.
This allows us to compare, to this reference case, results from following
sections where we explore the influence of key parameters.
In particular, we discuss the role of interior heating by solid accretion in
Section\,\ref{sec:resL0} (run \texttt{RADL0k0.01}).
We then proceed to investigate the role of the opacity in
Section\,\ref{sec:depop}, when accreting solids (run \texttt{RADL27k1}), or not
(run \texttt{RADL0k1}).
We leave exploring the role of the core mass itself to Section\,\ref{sec:depMc}
(runs \texttt{RADL27k1M15} and \texttt{RADL0k1M15}).

\subsection{Luminosity from accreting solids}

The structure of the envelope is set by the balance between the central
gravitational attraction of the planetary core and outward pressure from the
heated interior.
We now consider a planet where the interior heat generation is dominated by
energy deposition from solids sinking towards the core.
For such a planet, in the process of forming its 5\,M$_{\rm E}$-core, the
accretion luminosity is of the order
\begin{align}
  L_{\rm acc} &\approx \frac{GM_{\rm c} \dot M_{\rm s}}{r_{\rm c}}
  \nonumber \\
  &\approx 
  1.4 \times 10^{27}\,
  \left( \frac{\rho_{\rm c}}{3\,{\rm g/cm}^3} \right)^{1/3}
  \left( \frac{M_{\rm c}}{5\,{\rm M}_{\rm E}} \right)^{2/3}
  \left( \frac{\dot M_{\rm s}}{10\,{\rm M}_{\rm E}/{\rm Myr}} \right)
  {\rm erg/s}\,,
\end{align}
where $\dot M_{\rm s}$ is the solid accretion rate and $r_{\rm c}$ and
$\rho_{\rm c}$ the core radius and density.
Here, we have assumed most heat is ultimately deposited deep down in the
interior of the envelope \citep{Lambrechts_2014a}.

The magnitude of the solid accretion rate $\dot M_{\rm s}$ is poorly constrained. It
depends on the unknown amount of solids available in the
protoplanetary disc, but must be of the order of $10$\,M$_{\rm E}$/Myr
in order to grow the core to completion before disc dissipation. 
In Appendix\,\ref{app:pa}, we motivate the solid accretion rate in more detail. 

\subsection{Envelope structure}

\begin{figure}[t!]
  \centering
  \includegraphics[width=8.8cm]{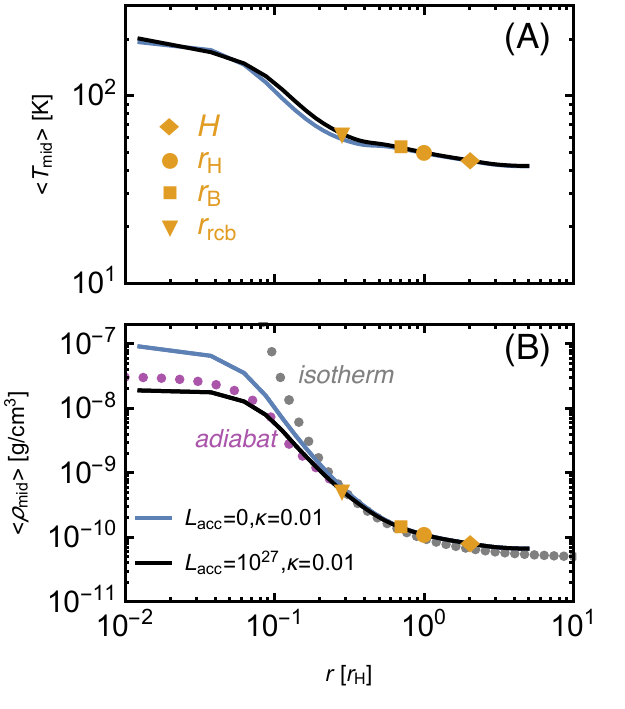}
  \caption{
	  Averaged envelope structure, for an envelope undergoing solid
	  accretion at a rate of 10\,M$_{\rm E}$/Myr (black) and an envelope
	  not undergoing solid accretion (blue).
	  Shown are the azimuthally averaged temperature and density
	  in the planetary midplane (\texttt{RADL27k0.01,RADL0k0.01}).
	  The diamond indicates the gas scale height $H$ of the protoplanetary
	  disc at the location of the planet, the circle the Hill radius
	  $r_{\rm H}$ of the $5$\,M$_{\rm E}$ core, the square the Bondi radius
	  $r_{\rm B}$, and the triangle the estimated position of the
	  radiative-convective boundary $r_{\rm rcb}$.  
	  The top panel (A) displays the temperature. 
	  The bottom panel (B) shows the azimuthally averaged density in the
	  midplane.
	  In the background we also show the density structure for a cooled-down
	  isothermal planet with a gray dotted curve. An adiabatic convective
	  interior, a polytrope with $\gamma=1.4$, is shown with the purple
	  dotted line up to $r_{\rm rcb}$.
  }
  \label{fig:avervar}
\end{figure}

\begin{figure}[t!]
  \centering
  \includegraphics[width=8.8cm]{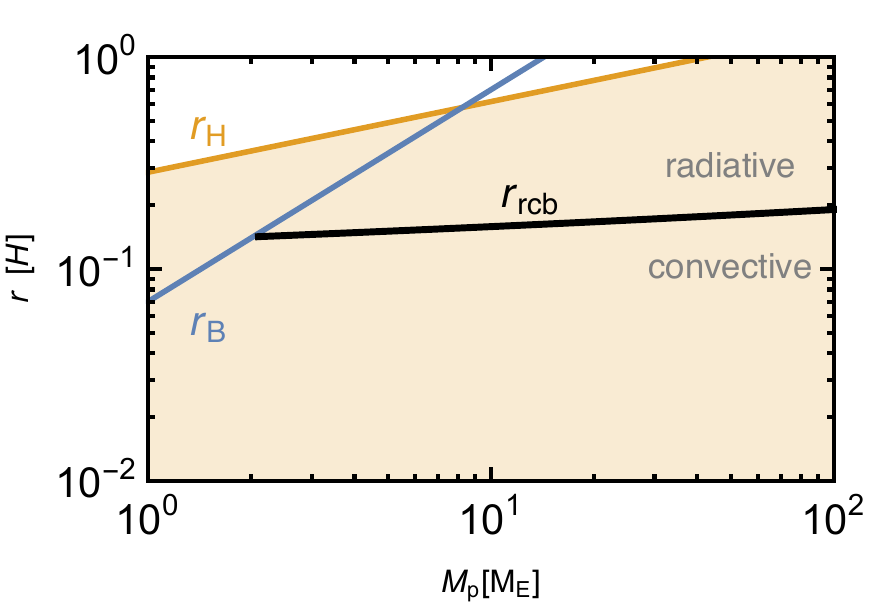}
  \caption{
	  Relevant length scales for a low-mass envelope, expressed with
	  respect to the disc scale height, as function of planetary mass. 
	  The yellow line represents the Hill radius ($r_{\rm
	  H}$), the blue line the Bondi radius ($r_{\rm B}$), and the black
	  curve the radius above which radiative energy transfer is
	  possible ($r_{\rm rcb}$), assuming $\kappa=0.01$\,cm$^2$/g and
	  $H/a_{\rm p} = 0.03$.
  }
  \label{fig:lscales}
\end{figure}

Fig.\,\ref{fig:avervar} shows the azimuthally-averaged temperature and density
profile of the envelope, together with characteristic length scales. 
We first note that gas is perturbed up to a scale height,
\begin{align}
  H = c_{\rm s} / \Omega_{\rm K}\,, 
\end{align}
away of the core.
Here $c_{\rm s}$ is assumed to be the unperturbed sound speed.
Therefore, gas feels the presence of the core even outside the gravitational
reach of the core, given by the  Hill radius,
\begin{align}
  r_{\rm H} = a_{\rm p} \left(\frac{M_{\rm p}}{3M_\odot} \right )^{1/3}\,,
  \label{eq:rh}
\end{align}
where $M_\odot$ is the stellar mass. Outside this radius, the stellar tidal
field starts dominating over the gravity of the core.
Since we consider low mass envelopes, we make the approximation that
the core dominates the potential ($M_{\rm p} \approx M_{\rm c}$).

The Bondi radius,
\begin{align}
  r_{\rm B} = \frac{GM_{\rm p}}{c_{\rm s}^2} \,,
\end{align}
corresponds to the radius where the pressure is perturbed by the
gravitational force from the planet, assuming no temperature perturbation.
This latter assumption makes this scale, often used in envelope studies, not
practical in the context of our radiative hydrodynamical calculations.
We therefore prefer scaling our results with respect to the Hill sphere.
However, for completeness we have indicated the Bondi radius with a square
symbol in Fig.\,\ref{fig:avervar}.

Interior to these outer length scales, spherical hydrostatic models predict the
existence of a radiative-convective boundary (Appendix \ref{ap:1D}).
This corresponds to the location in the envelope where heat transfer goes from
being convection-dominated in the interior to radiation-dominated in the outer
envelope.
The depth of this radiative shell, $r_{\rm rcb}$, is given by
\begin{align}
  \frac{r_{\rm rcb}}{r_{\rm B}} 
  &\approx \left[ \ln \left(
    \frac{64 \pi G \sigma_{\rm SB} \nabla_{\rm ad} T_{\rm mid}^4 M_{\rm c}}{3 \kappa L P_{\rm mid}} 
    \right) \right]^{-1} \,,
\end{align}
where $T_{\rm mid},P_{\rm mid}$ are the midplane temperature and pressure
and $ \nabla_{\rm ad}$ the adiabatic temperature gradient.
We discuss this expression in more detail in Appendix \ref{ap:1D}.

The radius of the core itself is located deep inside the envelope, at about
$10^{-3}$ Hill radii, and cannot be resolved.

In this paper, we consider low-mass planets embedded in the gas disc. 
Therefore the above length scales are typically ordered as $r_{\rm rcb} \lesssim
r_{\rm B} \lesssim r_{\rm H} \lesssim H$, as  can be seen in Fig.\,\ref{fig:avervar}.
The ordering does depend on the planetary mass and solid accretion rate.
Fig.\,\ref{fig:lscales} illustrates these various length scales as function of
planetary mass. 
We indicate the depth of the radiative zone with the black curve, assuming
a luminosity provided by the core accreting pebbles (see
Fig.\,\ref{fig:Lacc}, Appendix \ref{app:pa}).
Cores with masses above about $20$\,M$_{\rm E}$ are typically not longer
embedded in discs with aspect ratios below $H/a_{\rm p}=0.05$. 
Then the Bondi radius exceeds the Hill radius, which has become larger than the
local gas scale height of the disc. 

Taken together, the panels of Fig.\,\ref{fig:avervar} indicate a good
qualitative agreement with the expected thermodynamical structure of a planet,
based on 1 dimensional analytical calculations.
Overall, we note little temperature increase in the part of the envelope we
model (Panel A of Fig.\,\ref{fig:avervar}).
The outer region of the envelope remains nearly isothermal, while interior to
approximately $0.2$\,$r_{\rm H}$ the temperature slope steepens. 
This transition occurs around $r_{\rm rcb}$, which we calculated for the given
accretion luminosity.
This gives a first hint that heat transport in the interior of the planet does
not occur radiatively. 
This result is thus in line with standard 1D models.
Note that the turnover of the slope at the smallest scales towards
$0.01$\,$r_{\rm H}$ is largely an artefact of the necessary smoothing of the
potential.

Panel B of Fig.\,\ref{fig:avervar} shows the azimuthally-averaged density
structure in the envelope (black curve), as measured in the midplane.
We recover the characteristic exponential increase in the nearly isothermal
outer layer.
This can be seen by comparison to the gray dotted line in panel B of
Fig.\,\ref{fig:avervar}, which gives the density profile for a purely isothermal
envelope (an isotherm).
Interior to approximately $0.2$\,$r_{\rm H}$, corresponding to the location
of $r_{\rm rcb}$, we observe a turnover to a power law-like slope. 
This turnover qualitatively agrees with a polytropic interior, with adiabatic
index $\gamma=1.4$, which one would expect from an inner convective interior. 
This is illustrated by the purple dotted curve in Fig.\,\ref{fig:avervar}, where
we have taken into account the smoothing of the potential, which is responsible
for the flat slope at the smallest radii.

\subsection{Flow through the envelope}

\begin{figure}[t!]
  \centering
  \includegraphics[width=8.8cm]{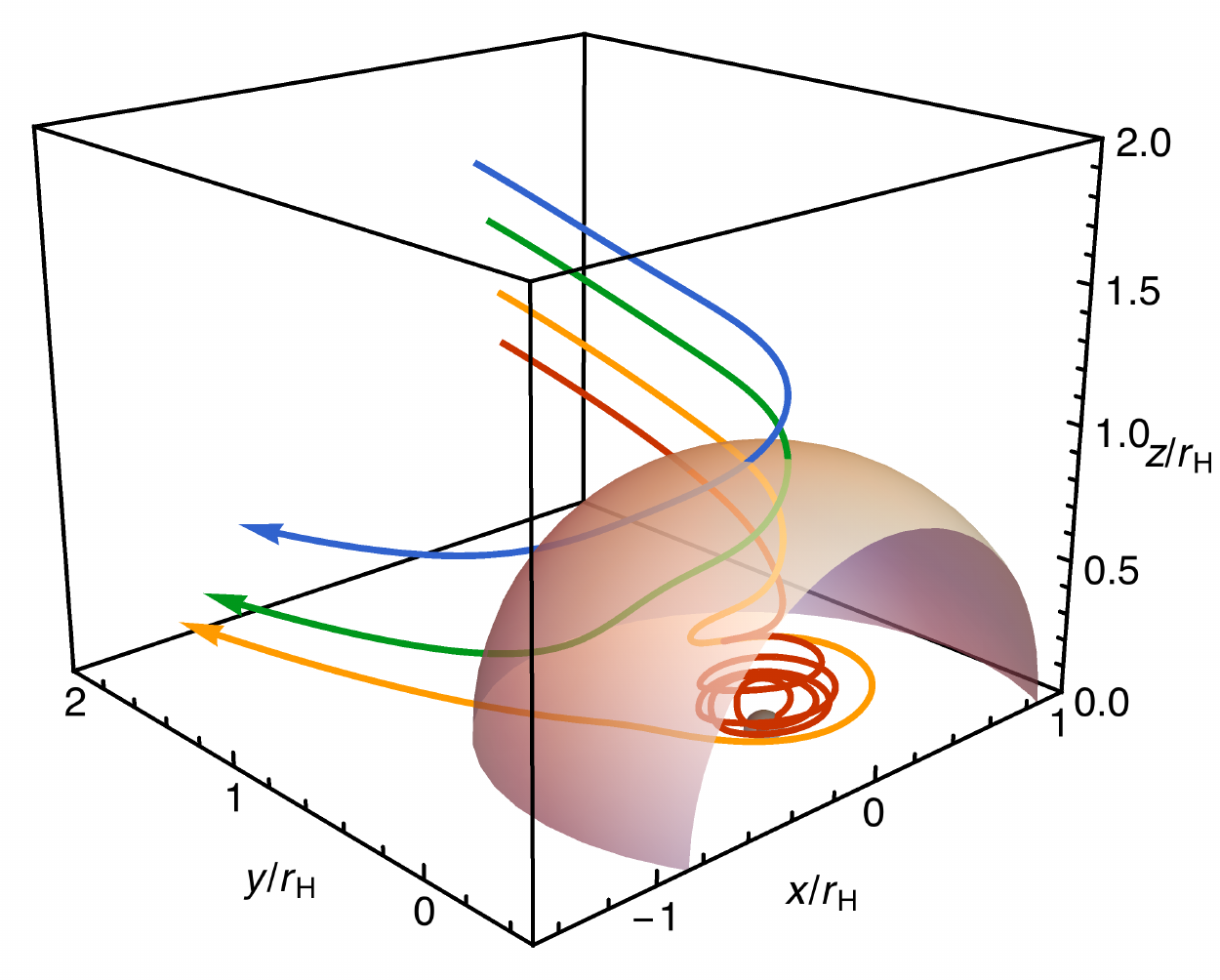}
  \caption{
	  Sample of gas streamlines,
	  arriving from the upstream direction with respect to the planet
	  (run \texttt{RADL27k0.01}).
	  The core is located at $x = r-a_{\rm p}=0$,  $y = a_{\rm p}
	  (\phi-\phi_{\rm p})=0$, $z=0$, as indicated by the black circle, and
	  surrounded by a transparent purple sphere with radius $r=r_{\rm H}$.
	  Results are shown in the frame co-rotating with the planet. 
	  Blue and green streamlines make a perturbed horseshoe orbit.
	  Gas parcels arrive well above the midplane ($z>r_{\rm H}$), enter the
	  Hill sphere and subsequently exit closer to the midplane ($z<r_{\rm
	  H}$).
	  The red and orange curves are examples of streamlines that interact
	  strongly with the core. 
	  The orange streamline circulates closely around the core (within
	  $r<0.5$\,$r_{\rm H}$), while the red streamline indicates a gas
	  parcel that remains trapped for the duration of the integration.
  }
  \label{fig:streamline}
\end{figure}

\begin{figure}[t!]
  \centering
  \includegraphics[width=8.8cm]{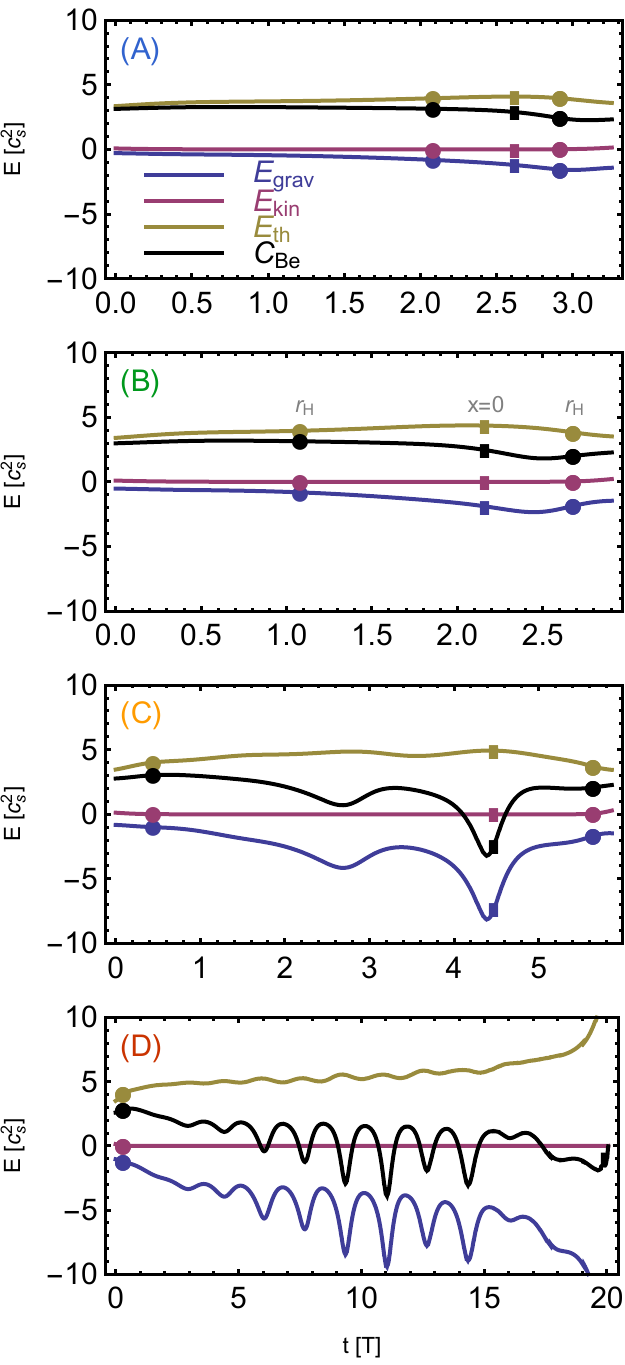}
  \caption{Energy balance along streamlines as function of time, in units of
  orbital period $T$.
  Each panel traces the energy budget for the different streamlines depicted in
  Fig.\,\ref{fig:streamline}, with panels A to D decreasing in height above the
  midplane. 
  Magenta curves show the kinetic energy contribution, which is negligible.
  The purple curve is the effective potential term (see main text for full
  description).
  The green curve gives the thermal energy along the streamline.
  The black curve is the sum of the energy contributions, the Bernoulli
  parameter.
  Circles mark the points where the field line enters and exits the Hill
  sphere. The square symbol indicates the point where the fluid parcel
  crosses the $x=0$ plane, which corresponds to the orbit of the planet. 
  When multiple crossings occur the point closest to the core is selected.
  }
   \label{fig:streamlinesEbalance}
\end{figure}

\subsubsection{Overview}

A large fraction of the gas inside the envelope, both in mass and volume, is not
bound to the core.
Instead, gas parcels enter and leave the planetary atmosphere in a complex
fashion on orbital timescales. 
We find these flows reach a (nearly) steady state pattern
and do not exceed the sound speed.

Fig.\,\ref{fig:streamline} illustrates the characteristic orbits of
gas parcels that enter the Hill sphere of the planetary core.
At high latitudes above the disc midplane, fluid elements are only weakly
perturbed by the planetary core, and perform the expected u-turn
associated with horseshoe orbits. 
However, at lower latitudes, streamlines get bend significantly when they
enter the Hill sphere and exit the atmosphere closer to the midplane (blue and
green curves in Fig.\ref{fig:streamline}). 
Closest to the core, we note that field lines (red curve in
Fig.\ref{fig:streamline}) remain trapped and perform a circumplanetary
motion with a weak retrograde tendency.

\subsubsection{Force balance}
We now inspect the force balance on the fluid elements that approach the core.
Assuming the Coriolis term dominates over advection (small Rossby number
approximation), the momentum equation in the rotating frame takes the form:
\begin{align}
  2 (\vc \Omega \times \vc u ) = - \frac{\nabla P}{\rho} - \nabla \Psi'\,.
  \label{eq:geobal}
\end{align}
Here we ignored the viscosity term and assumed steady state.
This expression is similar to geostrophic balance, but with the addition of an
effective potential term, which is the sum of the gravity terms and the
centrifugal force ($\nabla \Psi'$).
By multiplying Eq.\,\ref{eq:geobal} by $\rho$ and taking the curl, we find the form 
\begin{align}
  2 \left[ \nabla \times (\vc \Omega \times  \rho \vc u ) \right]
  &= 
  - \nabla \times (\rho \nabla \Psi' ) \, .
\end{align}
We can further rewrite this expression by expanding both terms
\begin{align}
  2 \left[ 
  - (\Omega \cdot \nabla ) (\rho \vc u )  
  + (\nabla \cdot \rho \vc u ) \vc \Omega
  \right]
  &= - \nabla \rho \times \nabla
  \Psi' - \rho \nabla \times \nabla \Psi'\, .
\end{align}
By making use of the steady state continuity equation, we can reduce this
expression to 
\begin{align}
  2 (\vc \Omega \cdot \nabla ) (\rho \vc u ) 
  &= \nabla \rho \times \nabla \Psi' \, .
  \label{eq:thermalwind}
\end{align}
Here, the directional derivative on the left hand term expresses the vertical gradients in the mass flux. 

Outside the Hill radius, densities are nearly uniform within a gas scale height
from the midplane. 
Therefore the above expression reduces to $\partial \vc u / \partial z =
0$, which corresponds to the Taylor-Proudman theorem. As $u_z = 0$ is zero in
the midplane, it remains so at higher altitudes as well.
Distant horseshoe orbits do not develop vertical motion
and appear approximately identical from the midplane up to higher altitudes. 

However, the planetary potential perturbs densities and the above approximation
breaks down near the planet.
From Eq.\,\ref{eq:thermalwind}, it is nevertheless clear that a vertical mass
flux gradient ($ \partial (\rho u_z) / \partial z $) requires special
non-aligned density gradients with respect to the gravity force term, which are
perpendicular to axis of rotation.
Such a situation occurs when horseshoe-like streamlines come close to the
planet and hit the spiral density perturbation triggered by the planet.
Indeed, we note a sudden drop in altitude towards the second quadrant
($x<0,y>0$) for the blue, green and orange streamlines of
Fig.\,\ref{fig:streamline}. 
As Eq.\,\ref{eq:thermalwind} is general in nature, similar vertical motions
are present in the isothermal approximation \citep{Fung_2015,Ormel_2015}.

We can also see that streamlines that get very close to the core and approach
the midplane, rotate in a retrograde fashion. This hints that the pressure term
slightly dominates the gravity terms towards the midplane, as can be seen from
Eq.\ref{eq:geobal}. 
Such resulting retrograde motion, as seen in our prograde rotating frame, is
reminiscent of vortices with high pressure centres in protoplanetary disc. 

Closer to the core we can expect hydrostatic balance to set in.
By definition this corresponds to the left hand terms of Eq.\,\ref{eq:geobal},
or equivalently the left hand term of Eq.\,\ref{eq:thermalwind}, being equal to zero. 
Assuming again no vertical velocities in the midplane, Eq.\,\ref{eq:thermalwind}
implies that there can be no significant vertical mass flux into an innermost
hydrostatic region. 

\subsubsection{Energy balance}

As fluid elements enter the envelope they convert part of the liberated
potential energy to thermal energy and exchange energy with the atmosphere. 
The energy budget on a streamline is given by
\begin{align}
  C_{\rm Be} = \frac{u^2}{2} + \Phi - \frac{(\vc \Omega \times \vc r)^2}{2} +
  \frac{P}{\rho} + e \,,
\end{align}
for steady flow in a non-time dependent potential, while ignoring heat exchange.
Here, the term $(\vc \Omega \times \vc r)^2/2$ comes from the centrifugal force
in the rotating frame and $e$ is the internal energy per unit mass. 
The last two term of the expression form together the enthalpy. 
If one assumes no heat exchange, and takes the inviscid limit, the quantity $
C_{\rm Be}$ is conserved along the field line (Bernoulli's constant). 
However, heat is exchanged between the disc and the planet, through
compressional heating and radiative exchange.

We carefully investigate the energy balance for the streamlines depicted in
Fig.\,\ref{fig:streamline}. 
This analysis is presented with one panel for each streamline of
Fig.\,\ref{fig:streamlinesEbalance}.
We first can note that the kinetic energy contribution is negligible for all
streamlines. 
This may look surprising given the short timescales, on the order of $10$
orbital timescales, on which fluid elements cross the Hill sphere, as one can
read from the horizontal time axis. 
However, from an energetic viewpoint, they can be safely ignored.
The potential term depicted here is the sum of stellar potential, the potential
of the planet and the centrifugal term $\Psi = \Psi_{\rm grav} - \Omega^2
r^2/2$. For clarity, in  Fig.\,\ref{fig:streamlinesEbalance} we only show the deviation from the background state at
the orbital radius of the planet ($\Psi_{\rm back} = -\Omega r - \Omega^2 r^2/2
= -1.5$ in code units).

As fluid elements approach the core, the combination of compressional heating
and irradiation increase the thermal energy ($E_{\rm th} = e + P/\rho$). 
The peak value occurs in the point closest to the planetary core.
The panels of  Fig.\,\ref{fig:streamlinesEbalance} shows that in general
internal energy increases as the potential component diminishes. 
However, this balance is not perfect, as one can see in the change in the
Bernoulli parameter during envelope crossing.  
Fluid parcels loose gravitational energy as they approach the point closest to
the core, put pick up energy on the way out of the envelope.
Overall, the tendency for streamlines is to produce a net deposit of energy,
as one can see by comparing the Bernoulli parameter at the entry and exit point
of the Hill sphere (dot symbols in Fig.\,\ref{fig:streamlinesEbalance}).

So far, we can conclude the following. 
The inspection of the averaged density and temperature structure
reveals a planet which appears to be in near hydrostatic equilibrium,
confirming  our standard picture of planetary envelopes. 
In contrast, the analysis of the streamlines paints a more complex picture.
There is no spherical symmetry. Moreover, streamlines reveal a strong steady
state flow through the planetary envelope, which explicitly relies on breaking
hydrostatic equilibrium. 
Little gas is bound to the envelope.
Additionally, the flow leaves an imprint on the energy transport in the
envelope, which we will investigate in more depth in
Section\,\ref{sec:heatcool}.

\begin{figure}[t!]
  \centering
  \includegraphics[width=8.8cm]{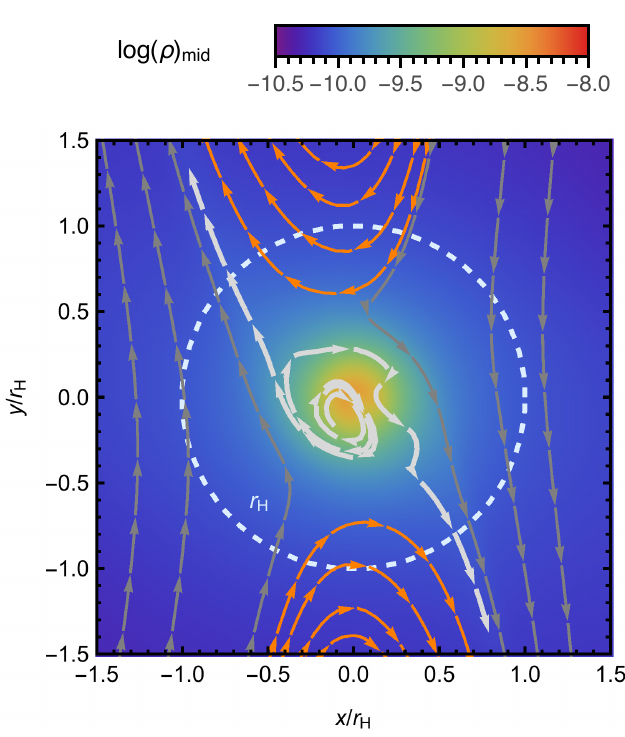}
  \caption{
	  Streamlines and gas density in the midplane of the planet, shown in a
	  co-rotating frame centered on a planet.
	  Here the core is accreting solids at rate of about 10\,M$_{\rm
	  E}$/Myr.
	  (\texttt{RUNHIGHRESL27}).
	  The x-axis points radially, the y-axis follows the azimuthal
	  direction.
	  Significant advection of mass occurs through the Hill sphere (marked
	  by the white dashed line).
	  The inner (left) and outer (right) circulating streamlines are shown
	  in gray.
	  Orange streamlines indicate the horseshoe region.
	  The white streamlines show infalling gas
	  from the poles being deflected outwards.
	  Inside this region gas is unsteady in nature but nearly bound. A slow
	  retrograde rotation can be seen, with respect to the rotating frame.
          The midplane gas density is shown in the background (in g/cm$^3$).
  }
  \label{fig:stream_midplane_rad}
\end{figure}

\subsubsection{Midplane}

For completeness, we also show the flow morphology in the midplane in
Fig.\,\ref{fig:stream_midplane_rad},
as is often done \citep{Tanigawa_2012, Angelo_2013, Fung_2015, Ormel_2015}.
Care has to be taken to not over interpret these midplane streamlines as being
a complete representation of the flow, because of the strong vertical flow
component that is present \citep{Tanigawa_2012}.
Additionally, because we forced mirror symmetry, we force $u_z=0$ in the inner
turbulent region where this condition does not necessarily needs to hold. 
In Fig.\,\ref{fig:stream_midplane_rad}, we recognize the horseshoe orbits,
depicted by the orange curves.
The horseshoe orbits are strongly asymmetric across the y-axis, compared to
standard 2D results \citep{Fung_2015}. 
Keplerian shear enters deeply within the Hill sphere, as indicated here by the
gray curves. 
The streamlines closest to the planet are unsteady close to the planetary core.
This is in line with the notion that the inner envelope is convective. 
Further from the planet we note the two channels in the midplane that are
aligned with the density wave perturbations launched by the planet, through
which flow can escape from the planet and the horseshoe region. 

Fig\,\ref{fig:stream_midplane_rad} can be compared to the velocity field in an
isothermal midplane, as depicted in Fig.\,\ref{fig:stream_midplane_iso}. 
The latter is characterized by two broad arms with flow escaping from the
planet, as depicted by the white curves, in line with \citet{Fung_2015} and
\citet{Ormel_2015}. In our non-isothermal calculation, the width of this two
arms is reduced, but the overall flow morphology is similar.

\subsection{Angular momentum}

\begin{figure}[t!]
  \centering
  \includegraphics[width=8.8cm]{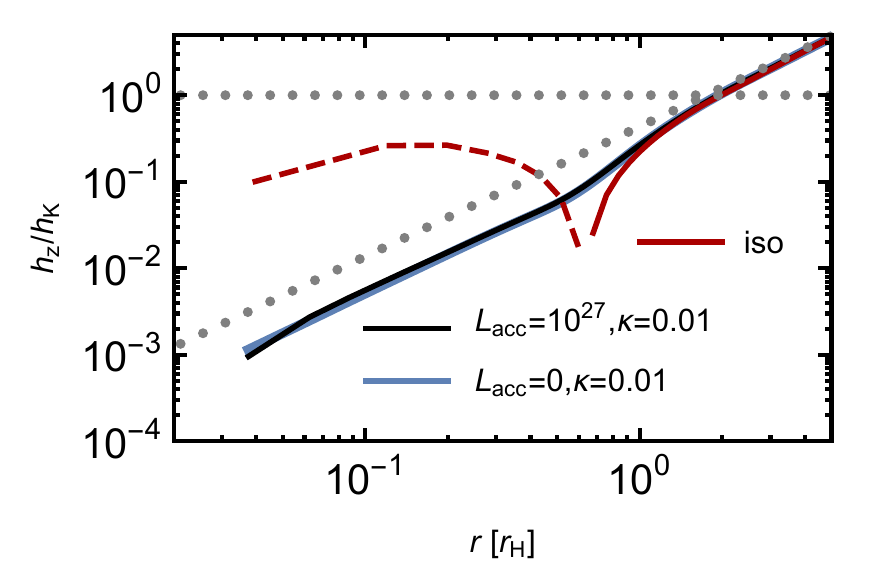}
  \caption{
  Azimuthally-averaged specific angular momentum, normalized by the Keplerian
  value, measured in the midplane.
  Full lines indicate retrograde motion, while dashed lines mark prograde
  motion. 
  The red curve shows an isothermal simulation (run \texttt{ISO}) with prograde
  rotation inside the Hill sphere. 
  The black and blue curve correspond to, respectively,  the reference
  simulations \texttt{RADL27k0.01} and \texttt{RADL0k0.01}.
  Gray dotted lines indicate two limit cases, gas purely in Keplerian rotation
  around the core ($h_{\rm z}/h_{\rm K} = 1$) and gas in purely Keplerian
  rotation around the star, as if no planet was present (Eq.\,\ref{eq:rrhhHill}).
  }
  \label{fig:rrhh}
\end{figure}

Isothermal simulations show the emergence of prograde discs around
planetary cores \citep{Machida_2008, Tanigawa_2012}. 
In our isothermal simulations we indeed recover prograde rotation
(Fig.\,\ref{fig:stream_midplane_iso}). 
Gas is orbiting close to Keplerian velocity, consistent with the results by
\citet{Tanigawa_2012} (their Figure 8).
In contrast, our non-isothermal simulations show the atmosphere does not
build up significant angular momentum. Only a weak retrograde motion is
measured, with respect to our prograde rotating frame.
This is illustrated in Fig.\,\ref{fig:rrhh}, which shows the azimuthally
averaged specific angular momentum, normalized by the Keplerian value,
\begin{align}
  \frac{h_{\rm z}}{h_{\rm K}} = \frac{\int_0^{2\pi} \rho r v_{\rm \phi}
  d\phi}{\int_0^{2\pi} \rho d\phi}  \frac{1}{\sqrt{GM_{\rm c} r} } \,,
\end{align}
for our reference simulations (runs \texttt{RADL27k0.01}, \texttt{RADL0k0.01},
\texttt{ISO}). Here, $r,\phi$ are radial and azimuthal coordinates centered on
the planet. 

We have thus found that envelopes are supported by a strong pressure force
compared to the gravitational force, as opposed to results in the isothermal
approximation.
This change in the radial force balance altered the rotation direction from
prograde to retrograde, as can be understood from Eq.\,\ref{eq:geobal}.
The planet thus only attracts low angular momentum accretion through the
planetary poles. 
We do not see the development of a flattened disc, as the thermal energy
contained in the envelope is not negligible compared to the gravitational
potential.
In fact, we find a result more consistent with Keplerian shear motion
persisting throughout the envelope.
In such a case, the angular momentum distribution takes the form,
\begin{align}
  \frac{h_{\rm z}}{h_{\rm K}}
    &\approx \frac{1}{2\pi}  \frac{1}{\sqrt{GM_{\rm c}}} r^{1/2}  
    \int_0^{2\pi} u_{\rm \phi} d\phi \,,
\end{align}
by ignoring azimuthal variations in the density. 
We can further linearise the velocity field to $u_y = -\frac{3}{2} \Omega_{\rm
K} x$, as in the shearing sheet approximation, to find
\begin{align}
  \frac{h_{\rm z}}{h_{\rm K}} 
  &\approx - \frac{3}{4}  \frac{1}{\sqrt{GM_{\rm c}}} \Omega_{\rm K}
  r^{3/2}\,.
\end{align}
Finally, by making use of the definition of the Hill radius  (Eq.\,\ref{eq:rh}),
we obtain
\begin{align}
 \frac{h_{\rm z}}{h_{\rm K}} &\approx
- \frac{3^{1/2}}{4} \left( \frac{r}{r_{\rm H}}\right)^{3/2} \,.
  \label{eq:rrhhHill}
\end{align}
This relation is shown in gray in Fig.\,\ref{fig:rrhh}. The scaling is
approximately maintained inside the Hill sphere, although averaged velocities do
become slower closer to the core.

\subsection{Mass flux through the envelope}

\begin{figure}[t!]
  \centering
  \includegraphics[width=8.8cm]{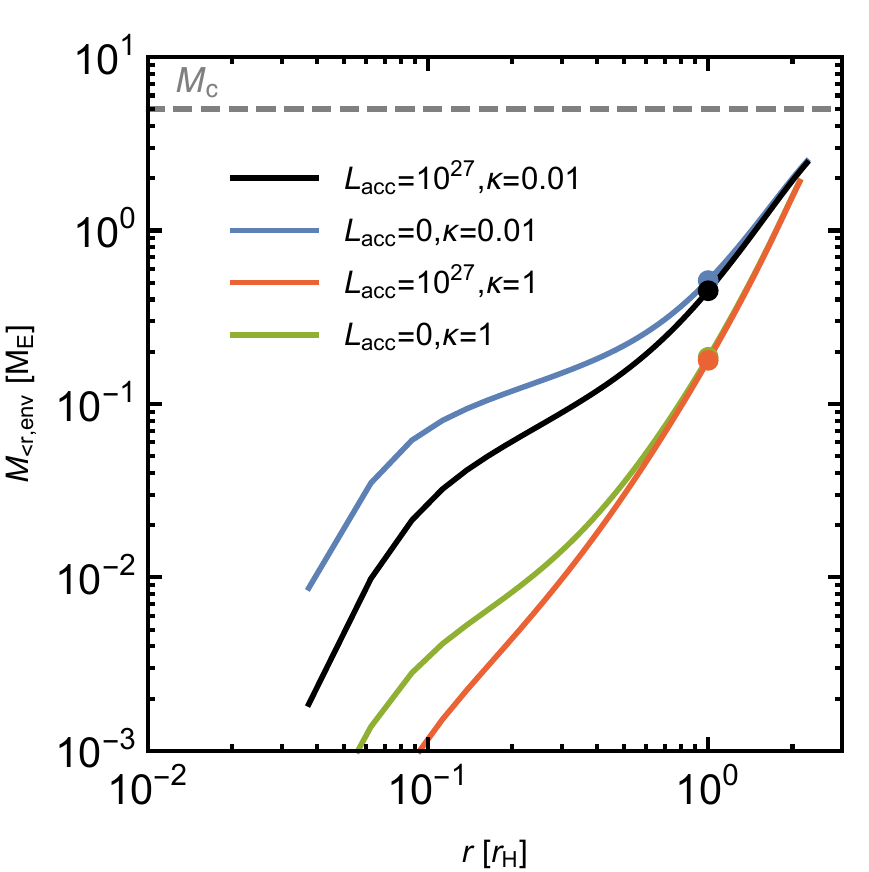}
  \caption{
	  Enclosed gas mass, as function of envelope radius $r$, around a
	  5\,M$_{\rm E}$-core (runs \texttt{RADL27k0.01}, \texttt{RADL0k0.01},
	  \texttt{RADL271}, \texttt{RADL0k1}).
	  The interior envelope mass goes down when solid accretion rates are
	  increased.
	  Higher dust opacities ($\kappa=1$cm$^2$/g) decrease the envelope mass.
	  The circles indicate the total interior mass within the Hill sphere.
  }
  \label{fig:rrMint5}
\end{figure}

\begin{figure}[t!]
  \centering
  \includegraphics[width=8.8cm]{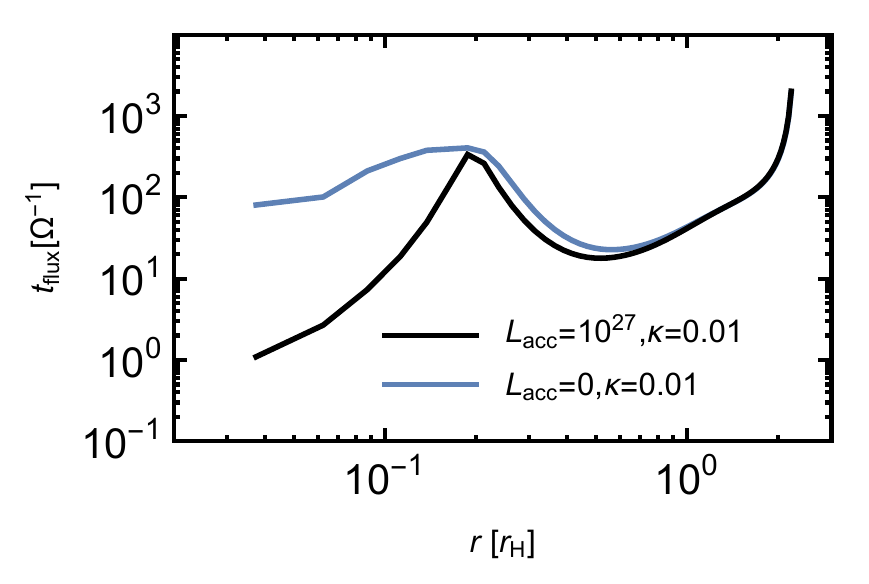}
  \caption{
	  Vertical mass flux through a shell with radius $r$.
	  The quantity shown is the timescale $t_{\rm flux}$, over which the
	  mass flux replenishes the  mass interior to a radius $r$.
	  The black curve corresponds to run \texttt{RADL27k0.01}, the blue
	  curve to run \texttt{RADL0k0.01}.
	  The total flow outside $r\gtrsim 0.3$\,$r_{\rm H}$ is dominated by 
	  vertical infall. 
	  The mass flow is of the order of a percent of the envelope per orbit.
	  For the planet accreting solids, $t_{\rm flux}$ decreases towards the
	  core because of convective motions that are not present in run
	  \texttt{RADL0k0.01}.
	  Near the core, interpretation is difficult, because of poor resolution
	  on these small scales.
  }
  \label{fig:rrtflux}
\end{figure}

The advection of gas is significant. 
Gas flows on tens of orbital timescales through the Hill sphere and involves a
significant mass fraction of the envelope. 
Fig.\,\ref{fig:rrMint5} shows how mass is distributed in the
envelope, by displaying the cumulative mass as function of the planetary radius. 
The total mass inside the Hill sphere is about $0.5$\,M$_{\rm E}$, about 10\,\%
of the core mass, as shown by the black curve in Fig.\,\ref{fig:rrMint5}. 
In our simulations we find the bulk of the envelope mass lays outside a radius
of $\approx0.2$\,r$_{\rm H}$.
The steep slope in the cumulative mass function at smaller radii
($\lesssim0.1$\,r$_{\rm H}$) is influenced by potential smoothing, as we could
also see in panel B of Fig.\,\ref{fig:avervar}.

We now introduce a vertical mass flux timescale, 
\begin{align}
  t_{\rm flux} = \frac{M(r)}{ \int_{S} \rho v_z ds } \,,
\end{align}
where $M(r)$ is the mass within a radius $r$, and the dominator is the
inwards directed vertical mass flux through the surface $S$ of the sphere. 
This is a mass-weighted measure of how rapidly gas is
transported through the envelope. 
It corresponds to the total vertical flow, which both includes gas pushed in and
again out of a spherical shell.
This quantity is displayed in Fig.\,\ref{fig:rrtflux}, as
function of envelope radius.

Outside $\approx0.2$\,r$_{\rm H}$, we note gas is pushed through the envelope
at a rate of about 
$10$\,\% of the envelope mass per orbit, 
up to a Hill radius from the core. 
These rates are consistent with previous measurements of isothermal simulations,
which are of the order of $1$\,\% of the envelope per $\Omega_{\rm p}^{-1}$
\citep{Ormel_2015} to
$100$\,\% per $\Omega_{\rm p}^{-1}$ \citep{Fung_2015}.
At larger radii, outside the Hill radius, the vertical flux diminishes as gas
comes in from directions perpendicular to to rotation axis. 

\begin{figure}[t!]
  \centering
   \includegraphics[width=8.8cm]{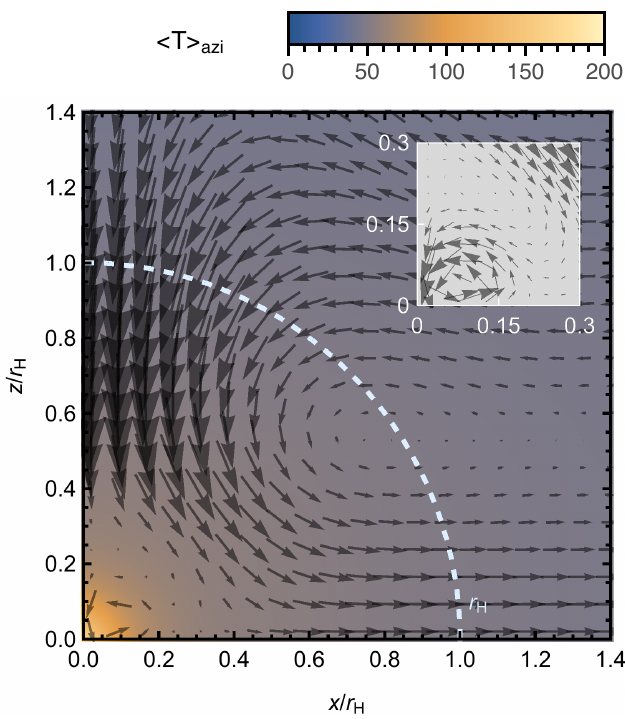}
   \caption{
	   Azimuthally-averaged velocity field around an accreting $5$\,M$_{\rm
	   E} $core, as seen in the co-rotating frame (run
	   \texttt{RADL27k0.01}).
	   The background shows the azimuthally-averaged 
	   temperature (in K)
	   around the core.
	   A large fraction of the gas that falls in from high elevation
	   through the poles gets compressed above the inner planetary
	   envelope, which gives rise to a mushroom-shaped  compressional heat
	   map.
	   The inset shows in more detail the azimuthally-averaged velocity
	   field, where overturning motions are driven by the heat release from
	   solid accretion.
	   }
  \label{fig:rzecompplotL27}
\end{figure}

Closer to the core, around $r \lesssim 0.2$\,r$_{\rm H}$, we note a peak in
$t_{\rm flux}$. 
It indicates that at this radius a large amount of the
vertical mass flux is deviated away from the core. 
This transition is illustrated in more detail by the azimuthally-averaged vector
field displayed in Fig.\,\ref{fig:rzecompplotL27}. 
This location broadly corresponds to the location of the radiative-convective
boundary. 

At even closer radii to the core, we do not measure an increase in
$t_{\rm flux}$, which one would expect from a clear transition to a static
bound envelope. 
Instead, an opposite trend is seen, where the vertical envelope mass flow
becomes more efficient. 
We associate this with the emergence of convective flows, as can be seen in
Fig.\,\ref{fig:rzecompplotL27}.
At close distances to the core ($r \lesssim 0.05$\,r$_{\rm H}$), the analysis is
complicated by poor resolution and future work will need to address the inner
envelope in more detail.

\subsection{Heating and cooling a 3-layer envelope}
\label{sec:heatcool}

\begin{figure}[t!]
  \centering
  \includegraphics[width=8.8cm]{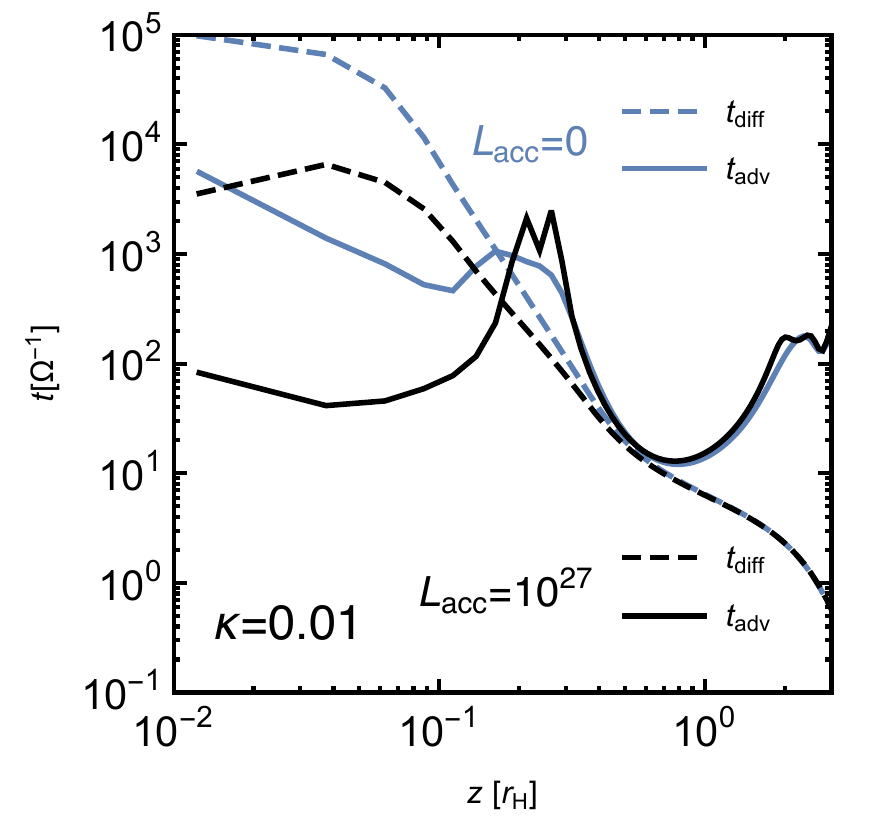}
  \caption{
	   Advection timescale (solid curves) and thermal diffusion timescale
	   (dashed curves) as function of the vertical depth in the envelope.
	   Black curves show results from the accreting planet
	   (run \texttt{RADL27k0.01}).
	   Outside the Hill radius, at heights above the scale height of the
	   disc, heat exchange by radiative diffusion is efficient. 
	   Inside the Hill Radius, around $0.5$\,$r_{\rm H}$, we find a balance
	   between advection and diffusion, as timescale become comparable and
	   are of the order of $10$--$100$\,$\Omega_{\rm p}^{-1}$. 
	   However, closer to the core, around $0.3$\,$r_{\rm H}$, we note that
	   the diffusion timescale is much shorter than the advection timescale,
	   which corresponds to a radiative zone in the envelope.
	   At even closer distances to the core, within $0.2$\,$r_{\rm H}$,
	   convective-like overturning motions become the dominant mode of
	   energy transport, as heat diffusion timescales reach timescales of
	   $10^4$\,$\Omega_{\rm p}^{-1}$.
	   Blue curves show results for the same planet, after the phase of
	   solid accretion (run \texttt{RADL0k0.01}). Advection and radiative
	   timescales follow a similar evolution with depth in the envelope. 
	   In the interior however, timescales increase by more than an order of
	   magnitude.
  }
  \label{fig:tdiff}
\end{figure}

The continuous flow of gas through the envelope also impacts heat transport.
The outer shells, where gas flow is dominant, cannot be seen as parts of a
static envelope that partake in secular cooling. 
Instead, they are dynamic regions, where a complex interplay of advection, heat
deposition by compression (vertical direction) and cooling (horizontal
direction) take place. 

In order to further investigate the modes of heat transport, we consider two
timescales: an advection timescale, 
\begin{align}
t_{\rm adv}  = \frac{r_{\rm H}}{u}\,,
\end{align}
and similarly a timescale to diffuse heat through radiation
over the characteristic length scale of the planetary envelope, 
\begin{align}
t_{\rm diff}  = \frac{r_{\rm H }^2}{D}\,.
\end{align}
Here $D$ is the radiative diffusion coefficient (see Eq.\,\ref{eq:Ddiff},
Sec.\,\ref{sec:Eeq}).

We find a 3-layer envelope structure. 
In the outer layer, between $0.5$ to $1\,r_{\rm H}$, advection of heat occurs on
a similar timescale as radiative diffusion, as can be seen in
Fig.\,\ref{fig:tdiff}.
Outside of the envelope, at $r\gtrsim 1\,r_{\rm H}$, radiative diffusion
starts to dominate as we move the heights above the disc midplane.

In a middle layer, between $0.2$ to $0.5\,r_{\rm H}$, we find that radiative
diffusion is the primary heat transport mechanism.
We can also identify this thin radiative shell in
Fig.\,\ref{fig:rzecompplotL27}, as region with little advection.

In the inner shell, within $r \lesssim 0.2\,r_{\rm H}$, advection
is again clearly the most efficient channel for heat transport.
This is consistent with an interior envelope, where convective motions
transport heat outwards.

\begin{figure}[t!]
  \centering
  \includegraphics[width=5cm]{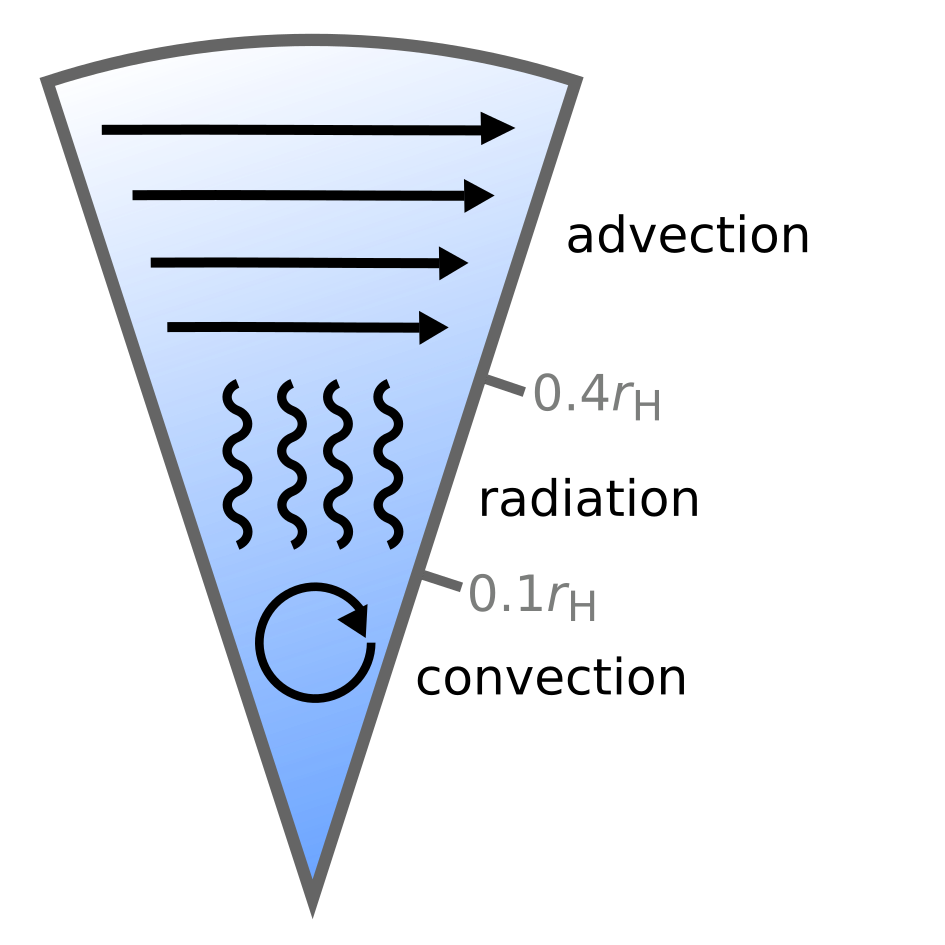}
  \caption{
  Cartoon illustrating a 3-layer envelope. 
  Energy transport occurs in the interior by convection up until 
  a radiative shell is reached, situated between $r \approx 0.1\,r_{\rm H}$ and
  $r \approx 0.4\,r_{\rm H}$. 
  Fast gas advection takes place in the outer envelope ($r \gtrsim 0.4\,r_{\rm
  H}$).
  }
  \label{fig:cartoon}
\end{figure}

\subsection{Summary}

So far, we have found that a planet in the process of accreting solids is
not in hydrostatic equilibrium, but does reach a steady state flow pattern. 
The envelope acquires a characteristic 3-layer structure, made up of a
convective interior, a radiative shell, and an advective outer envelope, 
which we illustrated with a cartoon in Fig.\,\ref{fig:cartoon}.
We observe that the gas flux remains substantial, even through
the deep interior, where the mass flux is on the order of 10\,\% of the
enclosed mass per orbit (Fig.\,\ref{fig:rrtflux}).
At larger radii, rapid advection of gas and radiative diffusion both help
transport heat away from the planet. 

The envelope around the solid accreting planet we described is stable and would
maintain steady state, were it not that on longer timescales the core mass
increases and solid accretion rates change.
When solid accretion dominates the luminosity budget, the cool-down process
of the planet can safely be ignored. 
Then planetary envelopes move from one pseudo-equilibrium state, similar to the
one described here, to the next. 
However, when solid accretion rates become too low, or come to a complete
standstill, this balance comes to an end.
In the next section, we describe an envelope in this latter stage.

\section{Results: after solid accretion}
\label{sec:resL0}

During the formation of a planet there is no guarantee solid accretion onto the
core continues during the whole life time of the gas disc.
Cores can get isolated from planetesimals due to scattering \citep{Tanaka_1999}
and resonant trapping \citep{Levison_2010}. %
Similarly, pebble accretion comes to a halt when cores grow sufficiently
massive to perturb the disc enough to from a pressure bump
that prevents inwards drift and accretion of pebbles \citep{Lambrechts_2014a}. 
Or pebble accretion simply comes to an end because radial drift depleted the
available solid reservoir before the onset of gas disc dissipation
\citep{Sato_2016}. 
Therefore, we now focus on the same planetary core, placed in the same disc
environment, but without interior heat generation by solid accretion (run
\texttt{RADL0k0.01}).

\subsection{Luminosity from compression}
\label{sec:luminosity}

\begin{figure}[t!]
  \centering
  \includegraphics[width=8.8cm]{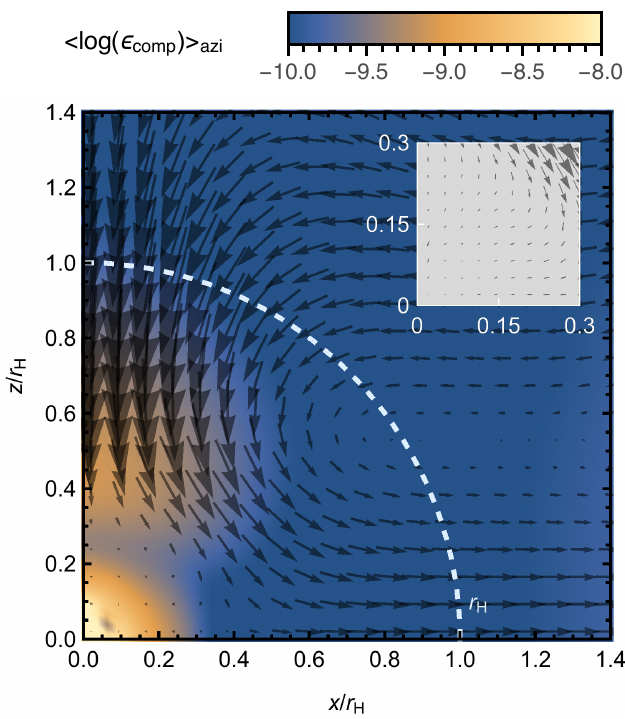}
  \caption{
  Azimuthally-averaged velocity field around a 5\,M$_{\rm E}$-core, similar
  to Fig.\,\ref{fig:rzecompplotL27}, but without the addition of accretion
  luminosity (run \texttt{RADL0k0.01}).
  The background shows the azimuthally-averaged compressional heating around
  the core.
  The inset zooms in on the vector field close to the core. 
  A small amount of convergent motion can be identified, which can be
  placed in contrast with the inset in Fig.\,\ref{fig:rzecompplotL27}, which
  shows convective overturn.
  }
  \label{fig:stream_aziaver_L0}
\end{figure}

\begin{figure}[t!]
  \centering
  \includegraphics[width=8.8cm]{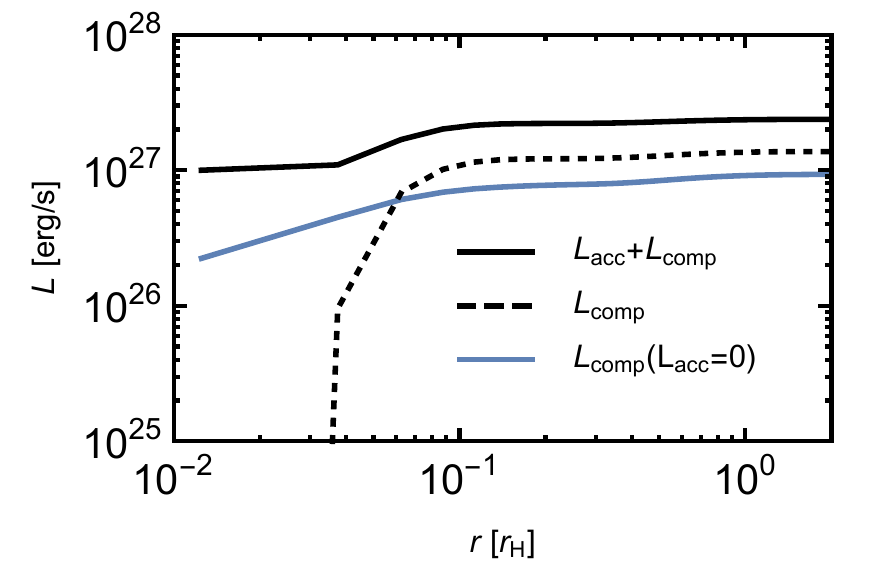}
  \caption{
	  Luminosity generated around the core. 
	  The blue curve shows the integrated energy released by compressional
	  heating as function of radius $r$, in the absence of heating by solid
	  accretion (run \texttt{RADL0k0.01}). 
	  Black curves correspond to run \texttt{RADL27k0.01}. The solid line
	  gives the sum of the accretion heat and the contribution by
	  compressional heating ($L_{\rm comp} = L_{\rm acc}+L_{\rm
	  comp}$). This latter fraction is depicted by the black,
	  dashed line.
  }
  \label{fig:rrLc}
\end{figure}

When the accretion of solids starts winding down, the dominant form of heat
generation becomes compressional heating of gas that enters the planetary
potential. 
Fig.\,\ref{fig:stream_aziaver_L0} shows the azimuthally-averaged compressional heating near the planetary centre and a heated cloud at the poles of the envelope where gas gets compressed as it gets deflected from the more bound interior towards the midplane.
The integrated net heating, as function of the planetary radius, is shown in
Fig.\,\ref{fig:rrLc}.
We find a luminosity of about $L \approx 10^{27}$\,erg/s, from compressional heating.

Note that we have obtained this new equilibrium envelope with corresponding
luminosity profile, without modelling the remnant interior heat in the envelope
left by the formation of the core and the secular cooling of the envelope.
We have thus modelled a cooled-down planet that has lost this heat contribution,
just before selfgravity becomes
important.
This latter stage in the evolution of the envelope is not accessible in
our simulations, as we de not model self-gravity.

Compressional heating is also relevant even when solid accretion occurs. 
Fig.\ref{fig:rrLc} shows compression of gas generates a similar amount of
luminosity as solid accretion. 
This energy is mostly liberated in the outer envelope.
Therefore, the location and total amount of energy deposition between the runs
with and without solid accretion are similar, but envelope structures are not
identical.

\subsection{Envelope structure}

The thermodynamical structure of the envelope appears similar to the planet in
the process of accretion solids, as can be seen in Fig.\,\ref{fig:avervar}.
The radial temperature profile is nearly identical. 
However, the density profile shows a sharper increase towards the centre,
compared to the solid-accreting case. 
This central density increase results in a slightly more massive inner
envelope, as can be inspected in Fig.\,\ref{fig:rrMint5}.

\begin{figure}[t!]
  \centering
  \includegraphics[width=8.8cm]{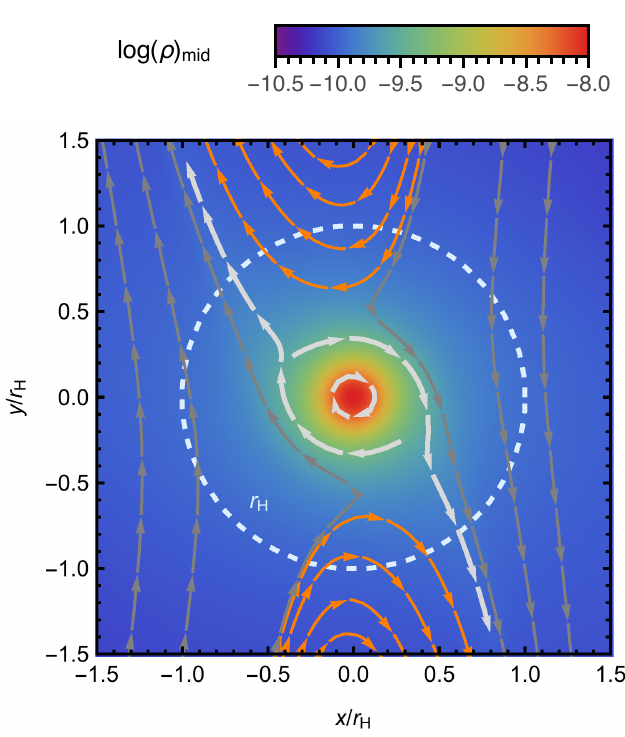}
  \caption{
	  Streamlines and gas density in the midplane of the planet, shown in a
	  co-rotating frame centered on a planet, not accreting any solids
	  (run \texttt{RUNHIGHRESL0}).
	  The x-axis points radially, the y-axis follows the azimuthal
	  direction.
	  Significant advection of mass occurs through the Hill sphere (marked
	  by the white dashed line).
	  The inner (left) and outer (right) circulating streamlines are shown
	  in gray.
	  Orange streamlines indicate the horseshoe region.
	  The white streamlines show infalling gas
	  from the poles being deflected outwards.
	  Inside this region gas is bound and rotates slowly retrograde.
  }
  \label{fig:stream_midplane_radK01}
\end{figure}

Inspection of the midplane streamlines, as displayed in
Fig.\,\ref{fig:stream_midplane_radK01}, also illustrates a similar flow
structure as in the solid-accretion phase (Fig.\,\ref{fig:stream_midplane_rad}). 
Rotation is again weakly retrograde, as also shown in Fig.\,\ref{fig:rrhh}. 
However the interior streamlines, do not reveal unsteady motion, instead we see
bound field lines in the midplane.

Indeed, the inner envelope shows little motion. The interior is in
near-hydrostatic balance and only a small amount of radial flow can be identified.
This can be seen in Fig.\,\ref{fig:tdiff}, the advection timescale becomes
exceedingly long as we approach the core. 
Nevertheless, the advection timescale is still shorter that the radiative
diffusion timescale in the inner optically thick envelope ($r\lesssim 2\,r_{\rm
H}$). 

\subsection{Advection of gas}

We again measure a strong mass flux through the outer envelope, of the order of
10\% of the envelope mass per orbit, similar to the case of the solid accreting
core, as shown in Fig.\ref{fig:rrtflux}. 
Closer to the core, within $r\lesssim 2\,r_{\rm H}$,  $t_{\rm flux}$, 
the less turbulent interior leads to a slower vertical mass flux timescale.
Previous studies have argued that this mass flux through the envelope can be estimated as a modified Bondi flow, 
\begin{align}
 \dot F_{\rm B} \sim \rho_{\rm mid}  u_{\rm char} r_{\rm B}^2 \, ,
\end{align}
where $u_{\rm char}$ is a characteristic velocity, often taken to be the
Keplerian shear and the cross section is typically taken to be the Bondi radius
 \citep{Angelo_2008, Lissauer_2009, Ormel_2015}. 
For the planet we consider, this corresponds to a mass flux of the order
$10^3$\,M$_{\rm E}$/Myr, which is indeed in reasonable agreement with our
measurement of the flow through the outer envelope (Fig.\ref{fig:rrtflux}).
Previous studies \citep{Angelo_2008, Lissauer_2009, Machida_2010} have
interpreted this mass flux as a measurement of the accretion rate, implying
unrealistic fast envelope growth timescales. 
Instead, we have show here that most of this mass flux is caused by an unbound
flow through the envelope.

\subsection{Accretion of gas}
\label{sec:L0acc}
A direct measure of the gas accretion rate, the net gas flux that remains bound
to the core, is difficult to make. 
The flow of gas that transits the envelope dominates the much smaller fraction
that is accreted.
Directly measuring the mass growth of the envelope is difficult as this occurs
on timescales that are numerically not feasible. 
In Appendix\,\ref{ap:res}, we show results of the longest integrations
which are numerically feasible, of the order of $50$\,orbits. 
A accurate measure of any direct mass growth of the planetary envelope cannot
be made.

We can however indirectly place a constraint on the gas accretion rate.  The
generation of heat requires a net divergence of the flow ($-P\nabla \cdot \vc u
$), which we can link to the accretion flow (inset of Fig.\,\ref{fig:stream_aziaver_L0}). 
As the luminosity generated by compressional heating is ultimately
the liberation of potential energy, we can estimate the gas accretion rate as
\begin{align}
  \dot M_{\rm gas}  &\sim \frac{r_{\rm char}}{GM_{\rm c}} L \, ,
  \label{eq:mdotgasestimate}
\end{align}
where $r_{\rm char}$ is a characteristic radius, corresponding to depth of
the potential. 
Taking $L \sim 5 \times 10^{26}$\,erg/s, and  $r_{\rm char} \sim r_{\rm
core}$ we would find an accretion rate of the order of about $\dot M_{\rm gas}
\sim 1$\,M$_{\rm E}$/Myr. 
This slow accretion estimate is in line with 1D-models \citep{Piso_2014}, and
further illustrates the challenge of making direct measurements of the
gas accretion rate.

\section{Dependency on opacity}
\label{sec:depop}

\subsection{opacity in protoplanetary discs}

Inside the planet-forming regions of protoplanetary discs, dust particles
smaller than mm in radius are the dominant cause of absorption and scattering
of radiation. 
The opacity caused by dust can be expressed as, 
\begin{align}
  \kappa_{\rm ice} \approx 5.8\,\frac{Z_{\rm dust}}{0.01}
  \left( \frac{T}{170\,{\rm K}}\right)^{2} \,{\rm cm}^2{\rm /g}\,,
\end{align}
just outside the ice line at approximately $170$\,K. 
Here, we have taken a composition and dust-to-gas ratio $Z_{\rm dust}$ similar
to the interstellar medium (ISM) from which the star and protoplanetary disc
formed.
Inside the ice line, icy particles sublimate and opacities rapidly settle to that provided by 
silicate grains,
\begin{align}
  \kappa_{\rm dust} \approx 1.5\,\frac{Z_{\rm dust}}{0.01}
  \left( \frac{T}{210\,{\rm K}}\right)^{1/2}\,{\rm cm}^2{\rm /g} \,.
\end{align}
A more complete description can be found in
\citet{Bell_1994,Angelo_2013,Semenov_2003}. 

These values can best be interpreted as upper limits. As dust grows by
collisions into larger particles, the ratio of small dust to gas, $Z_{\rm
dust}$ will decrease. 
Nevertheless, in special some regions in the disc, for example around ice
lines, the complex interplay between growth, fragmentation and sublimation may
lead again to higher values.

Inside the envelopes of planets, the evolution of the dust opacity is poorly constrained. 
As a first approximation, one could assume the envelope inherits the same
opacity as the disc, valid when coagulation proceeds more slowly then
advection and settling of dust inside the envelope.
Alternatively, one can estimate how the dust opacity may be reduced by grain
growth inside the planetary envelope
\citep{Movshovitz_2010,Ormel_2014,Mordasini_2014}. 
These models find atmospheres with ISM-like opacities in the outer
parts of the envelope, that decrease radially inwards to values around
$10^{-3}$\,cm$^2$/g.
Nevertheless, these models do not take into account bouncing and fragmentation
of grains, as well as the large gas advection flows through the atmosphere,
that will incorporate dust grains.

In the deep interiors of planetary envelopes, within the sublimation line for
dust $T_{\rm dust} \sim 1000$\,K, opacities are set by the gas component.
Such high temperatures are only reached very close to the core, within about
$10$ core radii \citep{Piso_2014, Lambrechts_2014a}. Our simulations do not
resolve such short distances. 
Therefore, we do not reach dust sublimation temperatures in the envelope, as
can also be seen in the top panel of Fig.\,\ref{fig:avervar}.
Consequently, we are not required to include gas opacities in our calculations.

Given these uncertainties in dust opacity values, we have limited this
study to opacities that are constant throughout the envelope. 
The previous results we showed were obtained for envelopes with $\kappa =
0.01$\,cm$^2$/g, which corresponds an envelope with moderate grain growth
\citep[as used in standard woks such as][]{Hubickyj_2005}.
We now will present envelopes where we assume no significant growth of dust has
occurred and opacities are close to an ISM-like opacity for silicate grains
($\kappa=1$\,cm$^2$/g). 
These unreduced opacities thus assume no significant dust depletion.
We will again consider two cases, the case of a planet accreting solids (run
\texttt{RADL27k1}), as well as the case where solid accretion has come to a
halt (run \texttt{RADL0k1}).

\begin{figure}[t!]
  \centering
  \includegraphics[width=8.8cm]{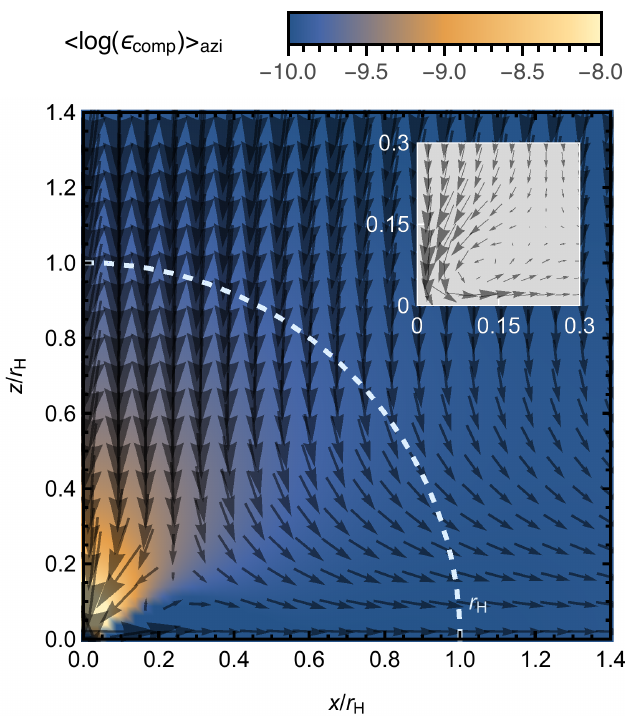}
  \caption{
	  Azimuthally-averaged velocity field around an accreting planet in a
	  high opacity environment (run \texttt{RADL27k1}). 
	  The gas flux from the poles enters even the deep interior, as is
	  clarified in more detail in the inset.
	  This stands in contrast to the low opacity case (run
	  \texttt{RADL27k0.01}, Fig.\,\ref{fig:rzecompplotL27}), where rapid
	  advection was limited to the outer envelope. 
  }
  \label{fig:rzecompplotL27k1}
\end{figure}

\subsection{Solid-accreting planet with unreduced opacity} 
\label{sec:sapho}

In a higher opacity environment ($\kappa=1$\,cm$^2$/g), gas flows through the
entire envelope of a solid-accreting planet ($L_{\rm
acc}=10^{27}$\,erg/s). 
This can be seen in Figure\,\ref{fig:rzecompplotL27k1}, which shows the
azimuthally-averaged velocity field and can be compared directly to
Figure\,\ref{fig:rzecompplotL27}.
Clearly, there is no trace left of the 3-layer envelope discussed in
Sec.\,\ref{sec:heatcool}. 
Instead gas circulates through the planet even at depths close to the core. 

\begin{figure}[t!]
  \centering
  \includegraphics[width=8.8cm]{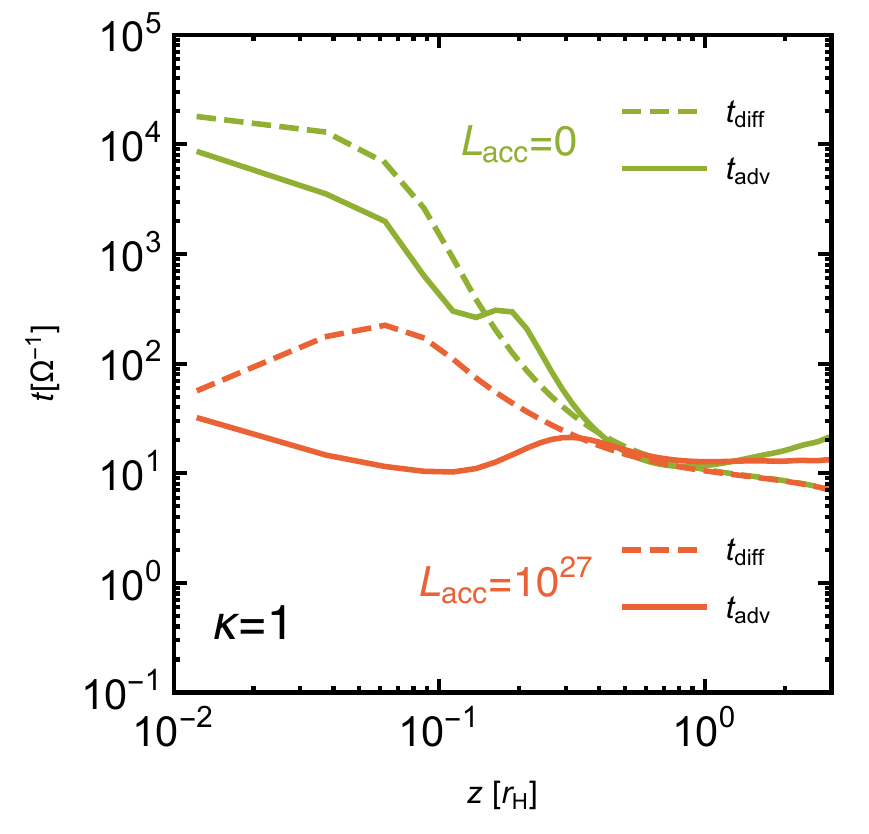}
  \caption{
	  Advection and thermal diffusion timescales through the envelope,
	  similar to Fig.\,\ref{fig:tdiff}, but for an unreduced opacity
	  $\kappa=1$\,cm$^2$/g.
	  Red curves show results from run \texttt{RADL27k1}, which is the
	  model of a solid-accreting core. 
	  Advection timescales are short, of the order of $10$\,$\Omega_{\rm
	  p}^{-1}$ throughout the envelope. 
	  The case without energy deposition by solid accretion (run
	  \texttt{RADL0k1}, green curves) shows advection and diffusion time
	  scales are of the same order and  peak towards the inner envelope
	  where they become as long as $10^4$\,$\Omega_{\rm p}^{-1}$. 
  }
  \label{fig:rrtdiffk1}
\end{figure}

There is thus no development of a more bound interior shielded by a thin
radiative shell where $t_{\rm adv} > t_{\rm diff}$, as was the
case in the reduced opacity models.
Instead, radiative diffusion timescales remain of the same order as the
advection timescales, as can be seen from the red curves in Fig.\,\ref{fig:rrtdiffk1}.
Radiative diffusion is overall much more efficient, compared to the
low opacity case, presented in Fig\,\ref{fig:tdiff} (run \texttt{RADL27k0.01}). 
It might appear counter-intuitive that higher opacities promote more efficient
radiative diffusion. 
However, the improved radiative transport is mostly just a consequence of the
lower central densities. Indeed, the envelope has a much smaller mass in the
interior, as can be seen in Fig.\,\ref{fig:rrMint5}.

It is worthwhile to reflect on this unexpected critical transition obtained by
simply increasing the opacity. 
A new equilibrium state emerged, where the envelope is dominated by the
advection of gas that enter from the poles.
This stands in contrast to spherical hydrostatic models which would have
predicted that increasing the opacity would result in the envelope becoming
simply more convection dominated. 
In such models the depth of the radiative zone is approximately inversely
proportional to the opacity ($P_{\rm rcb} \propto (\kappa L_{\rm acc})^{-1}$,
Eq.\,\ref{eq:PrcbPmid}). 
Indeed, we verified that this scaling approximately holds in the low
$\kappa$-regime by recovering equivalent envelopes between run
\texttt{RADL27k0.01} and a run with $\kappa=0.1$\,cm$^2$/g and $L_{\rm
acc}=10^{26}$\,erg/s, which holds the product $\kappa L_{\rm acc}$ constant.
However, as we have shown here, when the opacity is increased enough the
radiative outer zone starts to shrink up to point where a sudden transition
occurs. Then the envelope opens up and becomes advection-dominated.

Finally, we verified this result once more by analyzing a simulation with an
initial condition that was different from the standard procedure obtained by
gradually introducing the planet in the disc, as outlined in
Sec.\,\ref{sec:NumIm}.
Instead, the equilibrium state of \texttt{RADL0k1} was used as the initial
condition. After initialization, the envelope gradually got heated by accretion
heating and we found an equivalent envelope structure this way as well.
Therefore the fast advection result is not dependent on the method used to
initialise the simulation.

\subsection{After solid accretion, with unreduced opacity} 

\begin{figure}[t!]
  \centering
  \includegraphics[width=8.8cm]{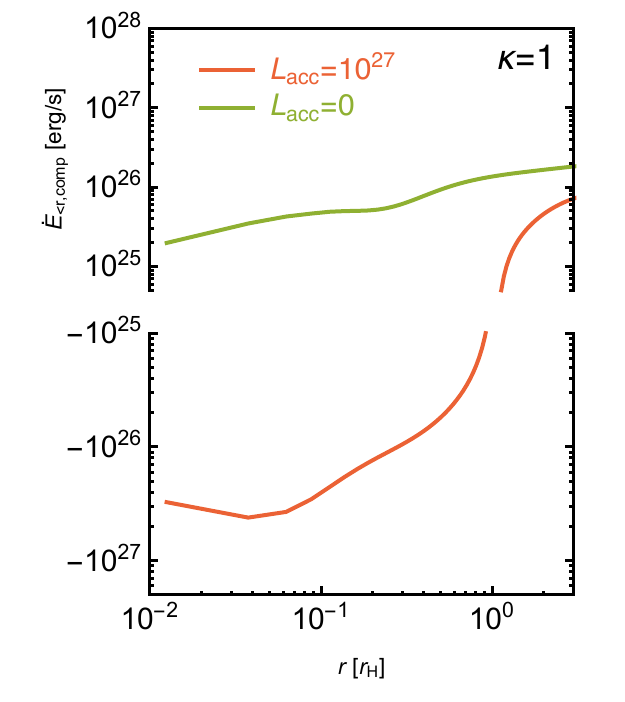}
  \caption{
	  Integrated energy delivery and removal by respectively compression and
	  expansion, for the unreduced opacity environment
	  ($\kappa=1$\,cm$^2$/g).
	  Red curves show the case where solids are accreted (run
	  \texttt{RADL27k1}). 
	  Inside the Hill sphere, we dominantly see that inner envelope mostly
	  acts as a sink of energy that helps gas parcels to expand as they
	  move through the envelope.
	  Green curves represent the energy deposition from compressional
	  heating after solid accretion (run \texttt{RADL0k1}).
	  Compressional heating now provides the luminosity that supports the
	  envelope. 
	  Note the y-axis has been cut to represent the negative values on the
	  log axis.
  }
  \label{fig:rrLck1}
\end{figure}

We now consider the case where solid accretion has come to a halt in an
unreduced opacity environment ($\kappa=1$\,cm$^2$/g), corresponding to run
\texttt{RADL0k1}.
We recover an envelope similar to the case with lower opacity (run
\texttt{RADL0K0.01}). 
Diffusion and advection time scales throughout the envelope are indeed similar,
as can be seen from comparing the diffusion and advection time scales (green
curves in Fig.\,\ref{fig:rrtdiffk1}) with those shown previously for run
\texttt{RADL0K0.01} in Fig.\,\ref{fig:tdiff}.

The luminosity released by compressional heating in the interior envelope is
approximately $L_{\rm comp} = 5 \times 10^{25}$\,erg/s, as shown by the green
curve in Fig.\ref{fig:rrLck1}. 
This is an order of magnitude lower compared to the case with
$\kappa=0.01$\,cm$^2$/g.
This reduction can be understood approximatively, because we recover a similar
depth of the radiative-convective boundary $r_{\rm rcb}/r_{\rm B}$, or
equivalently $P_{\rm rcb}/P_{\rm mid}$ when we force no accretion heat.
From Eq.\,\ref{eq:PrcbPmid}, we find $P_{\rm rcb}/P_{\rm mid}  \propto
\kappa^{-1}  L^{-1} P_{\rm mid}^{-1} T_{\rm mid}^4$ (for more
details see Appendix \ref{ap:1D}).
Therefore, one would expect $L \propto 1 / \kappa$ and a reduction of the
luminosity by a factor $100$. 
However, changing the opacity also changes the disc structure, and therefore the
values for the temperature and pressure. 
This softens the reduction in the luminosity to a factor $10$.
It also explains why the deposition $L_{\rm acc}=10^{27}$\,erg/s onto the same
$5$\,M$_{\rm E}$-planet with $k=1$\,cm$^2$/g is a significant contribution that
drastically changes the envelope structure, as discussed in
Sec.\,\ref{sec:sapho}. 
The interior of the envelope does not longer generate a supporting luminosity
from compressed gas, instead gas heats up and expands leading to a net cooling
effect, as can be seen from the red curve in Fig.\,\ref{fig:rrLck1}.

The lower luminosity we measure implies slower gas accretion rates. 
Repeating the order of magnitude analysis in Sec.\ref{sec:L0acc}, we would find
a gas accretion rate of 
$\dot M_{\rm gas} \sim 0.1$\,M$_{\rm E}$/Myr, for a luminosity of  $L \sim
5 \times 10^{25}$\,erg/s, and  $r_{\rm char} \sim r_{\rm
core}$.
This would lead to the formation of planets that have envelopes up to 
$\sim$$10$\% of the total planetary mass, similar to the known
Super-Earths and ice-giants.

\section{Dependency on core mass}
\label{sec:depMc}

The mass of the core is, besides the accretion rate and the opacity, an
important quantity that regulates the structure of the envelope. 
We have now seen that in environments where it is difficult for heat to diffuse
out of the envelope, the atmosphere can get advection-dominated.
We identified this effect around a low-mass core of $5$\,M$_{\rm E}$. 
Increasing the mass of the core should suppress this effect, as the potential
of the planet becomes too deep to allow the flow of gas to go through unhindered.

\begin{figure}[t!]
  \centering
  \includegraphics[width=8.8cm]{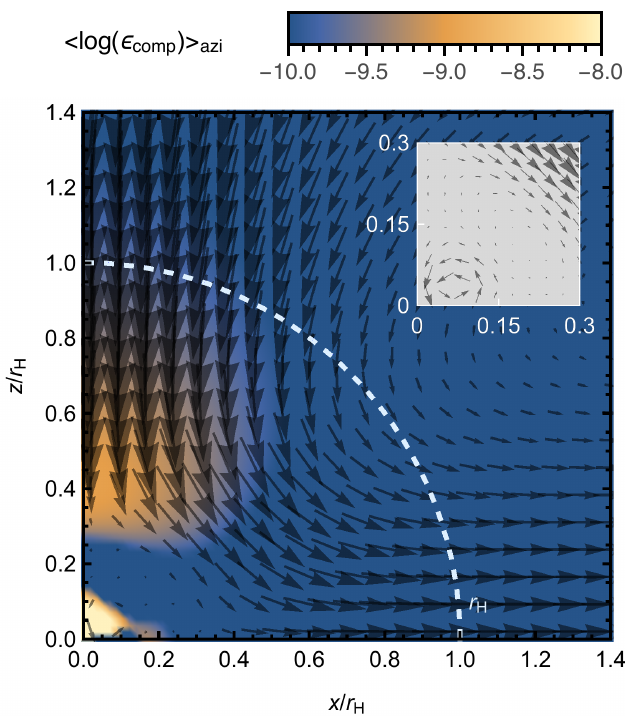}
  \caption{
	  Azimuthally-averaged velocity field around an accreting planet with a
	  $15$\,M$_{\rm E}$ in an unreduced opacity environment (run
	  \texttt{RADL27k1M15}). 
	  As opposed to the $5$\,M$_{\rm E}$-case, the bulk of gas that enters
	  through the poles does not longer enter the
	  deep interior. 
	  The deepened potential compensates for the higher opacity
	  ($\kappa=1$\,cm$^2$/g).
  }
  \label{fig:rzecompplotL27k1M15}
\end{figure}

Indeed, around a core of $15$\,M$_{\rm E}$ placed in the same unreduced opacity
environment with $\kappa=1$\,cm$^2$/g (run \texttt{RADL27k1M15}), we recover a
flow profile that shows great similarity to the one we observed around a
$5$\,M$_{\rm E}$-core, in low opacity environment. 
Fig.\,\ref{fig:rzecompplotL27k1M15} shows the azimuthally-averaged flow profile
around a core in the process of accreting solids. 
Clearly, we do not longer see the large scale advecting flows seen in
Fig.\,\ref{fig:rzecompplotL27} which occurred around the  $5$\,M$_{\rm E}$-core
in a similar $\kappa=1$\,cm$^2$/g environment.  

\begin{figure}[t!]
  \centering
  \includegraphics[width=8.8cm]{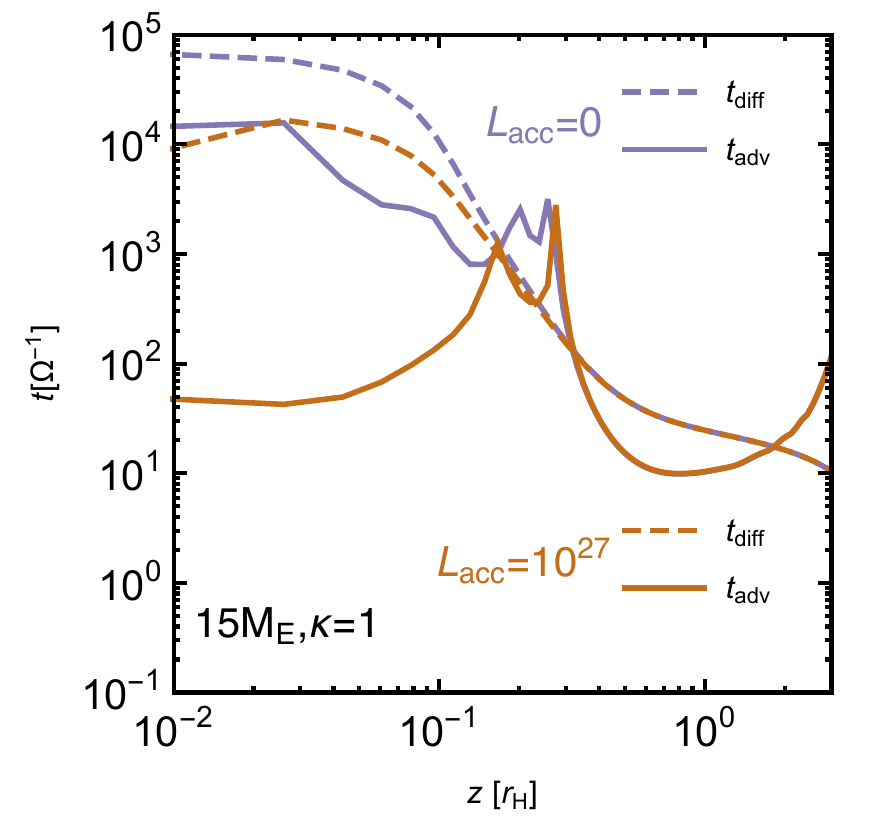}
  \caption{
	  Advection and thermal diffusion timescales through the envelope,
	  similar to Fig.\ref{fig:tdiff}, but measured now around a
	  $15$\,M$_{\rm E}$-core in an unreduced opacity environment with
	  $\kappa=1$\,cm$^2$/g (runs \texttt{RADL27k1M15} and
	  \texttt{RADL0k1M15}).
  }
  \label{fig:rrtdiffk1M15}
\end{figure}

The envelope recovers an exterior radiative-advective region, with interior a
shell where radiative transport dominates. Close to the center overturning
motions become the dominant mode for energy transport.
This can also be seen when inspecting the advection and diffusion timescales
presented in Fig.\,\ref{fig:rrtdiffk1M15}.

Also, without the contribution of solid accretion, we note that gas accretion
speeds up again (run \texttt{RADL0k1M15}). In this case, we measure a
luminosity from compressional heating of
approximately $5 \times 10^{26}$\,erg/s. 
Following the order of magnitude analysis of Eq.\,\ref{eq:mdotgasestimate}, we
find a gas accretion rate of around $1$\,M$_{\rm E}$/Myr, about an order of
magnitude larger than around the $5$ Earth mass core in a similar unreduced
opacity region. 
This facilities the transition to runaway gas accretion, but the growth rate is
still modest and disc dissipation might occur before growing a massive envelope.

\section{Implications: delaying runaway gas accretion}
\label{sec:Imp}

Previous studies \citep{Mizuno_1980, Pollack_1996, Ikoma_2000, Rafikov_2006,
Piso_2014} investigated the role of the core mass, opacity, solid accretion rate and envelope composition, but assumed hydrostatic envelopes. 
Here, we have seen that envelopes are not in hydrostatic balance, and large
flows of gas enter and leave the planetary atmosphere.
This result has several significant implications on the growth of the envelope.

\subsection{Dust opacity}

Previous studies have speculated that envelopes could acquire atmospheres up to
$100$ times enriched in opacity-providing grains, with respect to the star and
the disc \citep{Lee_2014,Venturini_2016}. 
Such high opacities would delay the contraction of the envelope.
However, this pathway can now be dismissed on three grounds.
High dust-to-gas ratios do not remain homogeneously mixed
\citep{Lambrechts_2016}, coagulation aids the rain out of solid from dust-rich
atmospheres \citep{Ormel_2014,Mordasini_2014}, and here we add that the rapid flux
of gas through the atmosphere does not allow for such dust pile ups in the first
place.
Conversely, extremely low dust opacities are also ruled out, because fresh dust is
continuously transported through the outer envelope.
Therefore, we expect dust opacities in the radiative-advective outer region of
planets to be, to a good approximation, comparable to the dust opacity in the
disc at the location of the planet. 
A more precise determination of the opacity at the radiative-convective
boundary would require to solve self-consistently for the transport of dust
particles and the opacity dependent depth of the advective outer shell.

\subsection{Accretion of solids and envelope pollution}

We expect the gas flows through the envelope to only have a moderate effect on
the accretion rate of solids.
Large planetesimals feel little gas drag, because of their large size.
Smaller pebbles on the other hand could see their trajectories changed.
However, accretion of pebbles dominantly occurs close to the disc midplane, from a
direction diagonal to the direction of outflow along the spiral density waves
\citep{Lambrechts_2012}.
Additionally, pebbles have friction and crossing times that are comparable, of
the order of the orbital timescale, while the flux through the atmosphere occurs
on 10 times as long timescales in the outer envelope.
Nevertheless, we plan to study this in more detail in an upcoming work, which
will also allow us to study at which depths in the envelope the accretion heat
is released. 

As solids sink in the envelope they can evaporate and pollute the envelope.
Particles with a silicate composition only vaporize at high temperatures
exceeding $1000$\,K that are found close to the core.
Therefore, these dust species settle in the inner nearly-hydrostatic
envelope and will be accreted.
Ices on the other hand can deposit water vapour at much shallower depths in the envelope.
If the sublimation point is located in the highly advective outer regions
of the atmosphere, vapour will simply be transported away in tens of orbital
timescales.
However, in our simulations where the planet is located outside of the
water ice line of the disc, we can see that this is not the case. 
In fact, the water ice sublimation point, at about $170$\,K, is located deep
inside the convective interior, shielded from the outer layers where rapid gas
advection takes place (Fig.\,\ref{fig:avervar}).
Therefore, not all water vapour will be immediately lost. 
Instead, the solid accretion rate, the redistribution of vapour by convection and the advection rate will set the fraction of solids
that settle to the core, the fraction that is deposited in the envelope and the
vapour fraction that is lost.
It will be important to quantify this in more detail, because this balance
influences core growth and the accretion of gas as well, since the deposition
of high mean molecular weight material speeds up gas accretion
\citep{Hori_2011, Lambrechts_2014a, Venturini_2016}.

\subsection{Cooling the envelope}
Finally, there is the impact on the secular cooling of the envelope. 
Gas in spherical hydrostatic models is bound. 
Therefore the outwards transport of heat is paired with cooling the envelope and
increasing the envelope mass.
In this way, a hydrostatic growth sequence can be constructed for planets
\citep{Pollack_1996,Ikoma_2000,Piso_2014}. 

Rapid advection of gas through the envelope challenges this paradigm
\citep{Ormel_2014,Fung_2015}.
The outward transport of heat is not extracted from a bound reservoir of gas,
which slows down envelope cooling and associated mass growth compared to
spherical hydrostatic models.
However, the picture is complicated.
After the solid-accretion phase, we have shown that planets have their
strong advecting flows limited to the outer envelope (Sec.\,\ref{sec:resL0}).
In the inner envelope, we see mass advection diminish as we get closer to the core.
Therefore, gas close to the core is in approximate hydrostatic balance and
modified 1D prescriptions may be suitable, allowing an approach as
in \citet{Angelo_2013}.

A more quantitative analysis determining the bound envelope will
require an improved resolution in the inner shells. Resolving this region close
to the core would also necessitate the use of a more accurate, non-constant,
opacity prescription that takes the high temperatures in the inner envelope into
account. An improved equation of state may also be necessary.
Such improved measurements
would aid in making secular cooling histories of young planets more accurate.
Furthermore, it would be of great value to explore also a larger planetary mass
range, since even planets as large as Jupiter see a large gas flux transit
through their atmospheres \citep{Gressel_2013,Morbidelli_2014,Szul_2014}.

\subsection{Summary}
In summary, we have identified three different effects that stall envelope
accretion. 
Dust opacity values will be similar to the disc, water pollution can be
inefficient, and cooling times are delayed by the delivery of fresh gas.

\section{Interpreting exoplanet Super-Earths and gas giants}
\label{sec:occ}

Why do some planets build up massive envelopes and others do not?
We have argued that the formation of gas giants with large atmospheres requires
specific conditions:
the rapid formation of cores larger than 15\,M$_{\rm E}$ in
a disc environment with reduced opacities.
These requirements can be met in the outer regions of the protoplanetary
disc, where particles have grown to cm-sizes and radially drift inwards through
the disc.
Such rocky/icy pebbles can be efficiently accreted by planetary embryos
\citep{Ormel_2010,Lambrechts_2012},
allowing for approximate 20\,M$_{\rm E}$-cores to form outside of the iceline
within disc lifetimes \citep{Lambrechts_2014a}.
Such fast core growth does generally require initial dust-to-gas ratios in the
disc that are solar-like or higher \citep{Lambrechts_2014b,Bitsch_2015}.
This requirement seems to be in line with the observed lack of close-in gas
giant planets around stars with sub-solar metallicities \citep{Buchhave_2012}.
We note that the required solar or higher initial dust-to-gas ratio does not
imply that densities of opacity-providing grains must have been high as well.
Particle growth outside the ice line reduces the opacity and local pebble
surface densities fall rapidly by a factor $10$ below initial values because of
radial drift \citep{Lambrechts_2014b}.

Super-Earths and ice giants did not accrete such massive gas envelopes.
Their smaller core masses helped prevent rapid gas accretion.
Additionally, they may have formed dominantly in regions where disc opacities were not significantly reduced.

Ice giants have cores that formed slowly in very wide orbits and likely never
experienced strongly reduced opacities, as dominant particle sizes remain small
in the outer disc \citep{Brauer_2008}.

Super-Earths on the other hand are commonly found in short orbits. They thus
formed in situ or the cores migrated into this inner disc region.
Possibly, they dominantly formed in a part of the disc where solids piled up,
which would explain why Super-Earths multiples are common \citep{Winn_2015} and
occurrence rates are only weakly dependent on stellar metallicity
\citep{Buchhave_2012}.
Such a pile-up of solids would likely have been associated with a high opacity
environment, because of dust delivery by radial drift and possibly collisional
fragmentation, which would further limit gas accretion.

In summary, we propose that gas giant planets dominantly form outside of the ice
line, where particles grew to pebble sizes, allowing the formation of large
cores. 
Models of exoplanet compositions support that gas giants typically
have substantial cores exceeding $10$\,M$_{\rm E}$ \citep{Thorngren_2016}.
Reduced opacities may then have further facilitated rapid gas accretion.
Super-Earths on the other hand have smaller cores, and may additionally have
formed in high opacity environments associated with pile-ups of solids in the
inner disc. This prevented rapid gas accretion.
The specific conditions for the formation of massive gas envelopes may explain in part why occurrence rates for gas giants are below Super-Earths by about a factor five \citep{Winn_2015}.

\section{Summary}
\label{sec:Sum}

Atmospheres around growing planets in the Super-Earth to ice giant regime are
not spherically symmetric envelopes in hydrostatic balance.
Here, we presented global 3D radiative hydrodynamical simulations that show a
steady state flow through planetary envelopes.
Large amounts of gas enter through the poles of the envelopes and exit near the
disc midplane. 
This is the result of density perturbations with gradients that are not aligned
with the gravitational potential. 

Our results imply that 
(1) opacities in the outer envelope are similar to disc opacities, 
(2) icy solids will sublimate in the convective interior and
(3) the flow of gas delays the secular cool-down of the envelope.
We numerically investigated the dependency of our results on three key
parameters; the accretion rate, the dust opacity and the mass of the core.

Dust opacities can fluctuate widely in the protoplanetary disc, because of
particle growth and drift.
In regions of the disc with dust opacities of $\kappa=0.01$\,cm$^2$/g, below
the initial ISM-like opacity of about $\kappa\approx 1$\,cm$^2$/g, planets
develop a 3-layer atmosphere.
The envelope consists of an advective outer shell, a radiative shell around a
radius of $0.3$ Hill radii, and an interior which shows convection-like
overturning motions that transfer the heat of solid accretion.
When solid accretion comes to a halt, gas close to the core slowly settles in
the potential. The liberated heat from this convergent motion supports the
envelope. 
Measurements of the accretion rate are challenging, but our estimates are of
the order of $1$\,M$_{\rm E}$ per Myr. 
Improved measurements will require long-term simulations exceeding the
advection timescale which are performed at high resolution in order to allow
smoothing lengths down to the core surface.

We also explored unreduced dust opacities that are comparable to values found
in the ISM ($\kappa=1$\,cm$^2$/g) in the disc and envelope, motivated by the
the high gas advection rates in the outer atmosphere. 
For planets in the process of accreting solids, we identify a previously
unreported transition into an advection-dominated envelope. 
Hydrostatic models would have predicted a nearly completely convective
atmosphere. 
Instead, we found an envelope where advection of gas, from the poles to the
deep interior, everywhere becomes the dominant mode of energy transport. 
After solid accretion comes to an end, gas accretion rates are low, of the
order of $0.1$\,M$_{\rm E}$ per Myr.
Under these conditions, a $5$-M$_{\rm E}$ core would not go into
runaway gas accretion and therefore remain a Super-Earth.

However, larger cores lead to more bound interior envelopes and an increase in
the gas accretion rate.
We found that around massive solid-accreting cores ($M_{\rm c} \gtrsim
15$\,M$_{\rm E}$) envelopes are no longer advection-dominated in a
$\kappa=1$\,cm$^2$/g environment.
After solid accretion, the increase in the core mass boosts the envelope growth
rate by a factor $10$, compared to the $5$\,M$_{\rm E}$ case. Therefore, a
transition to runaway gas accretion becomes possible again.

These findings deserve further attention and future models of envelope evolution
would benefit from a more quantitative analysis.
However, based on our current results we can already argue that Super-Earths
did not accrete massive gas envelopes because their cores are sufficiently
small, provided they formed in disc environments where dust opacities
were not strongly reduced below $\kappa \sim 0.01$\,cm$^2$/g.

\begin{acknowledgements}
We would like to thank 
Aur\'elien Crida and 
Alessandro Morbidelli
for insightful comments. 
We also benefited from helfpful discussions with
Bertram Bitsch,
Matth\"aus Schulik,
Julia Venturini,
Seth Jacobson,
Anders Johansen and 
Tristan Guillot.
The authors are grateful for the constructive feedback by an anonymous
referee.
We are thankful to ANR for supporting the MOJO project
(ANR-13-BS05-0003-01).
This work was performed using HPC resources from GENCI [IDRIS] (Grant 2016,
[i2016047233]) and from ``Mesocentre SIGAMM", hosted by the Observatoire de la
C\^ote d'Azur.
\end{acknowledgements}

\appendix

\section{Methods: Convergence tests}
\label{ap:res}

\begin{figure}[t!]
  \centering
  \includegraphics[width=8.8cm]{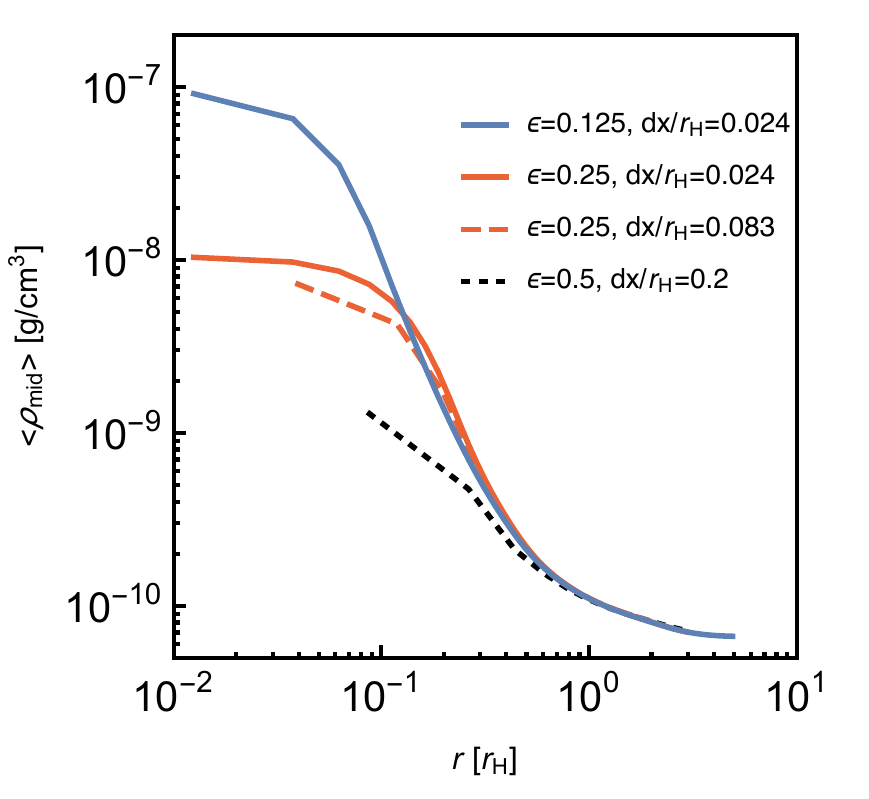}
  \caption{
	  Azimuthally-averaged density profile in the midplane, for different
	  smoothing lengths and resolutions.
	  The legend indicates the used smoothing length parameter $\epsilon =
	  r_{\rm sm}/r_{\rm H}$ and the characteristic grid cell width
	  inside the envelope ($dx/r_{\rm H}$). 
	  Only in our highest resolution runs, with a small smoothing length, 
	  do we correctly capture the envelope structure to within a radius of
	  $0.1$\,$r_{\rm H}$.
	  Inside this radius, one can start to see the influence of the
	  smoothing length, which reduces gas densities towards the core.
  }
  \label{fig:smooth}
\end{figure}

We model the presence of a planetary core through an imposed gravitational
potential for the gas. 
Figure \ref{fig:smooth} illustrates the dependency of our results on the
resolution across the Hill sphere and the employed smoothing length.
The planetary potential is given by 
\begin{align}
\Phi_{\rm p} = -\frac{GM_{\rm c}}{r}
\end{align} outside a smoothing
radius $r_{\rm sm}$ away form the planet.
Within the smoothing length, the central singularity is avoided
by a cubic interpolation to $\Phi_{\rm p} =-2GM_{\rm c}/r_{\rm sm}$, using
\begin{align}
   \Phi_{\rm p, sm} =
   -\frac{GM_{\rm core}}{r_{\rm sm}} 
   \left[
   2 
   - 2 \left( \frac{r}{r_{\rm sm}} \right)^2
   + \left( \frac{r}{r_{\rm sm}} \right)^3
    \right] \,,
\end{align}
following \citet{Klahr_2006}.
The mass of the envelope is ignored in the calculation of the potential, which
is a valid approximation for the low-mass envelopes we consider.

The potential is centered in between grid cells. 
We found the planetary potential to be well resolved with about $\gtrsim 4$
grid cells across a smoothing length. 
This avoids additional artificial smoothing of the potential,
which can be seen in the low resolution results in Fig.\,\ref{fig:smooth}. 
Finally, we introduce the potential gently over an orbital period $T$, as 
\begin{align}
  M = M_{\rm c} \left[\sin \left( \frac{\pi}{2} \frac{t}{T} \right) \right]^2 \, .
  \label{eq:intropot}
\end{align}

\begin{figure}[t!]
  \centering
  \includegraphics[width=8.8cm]{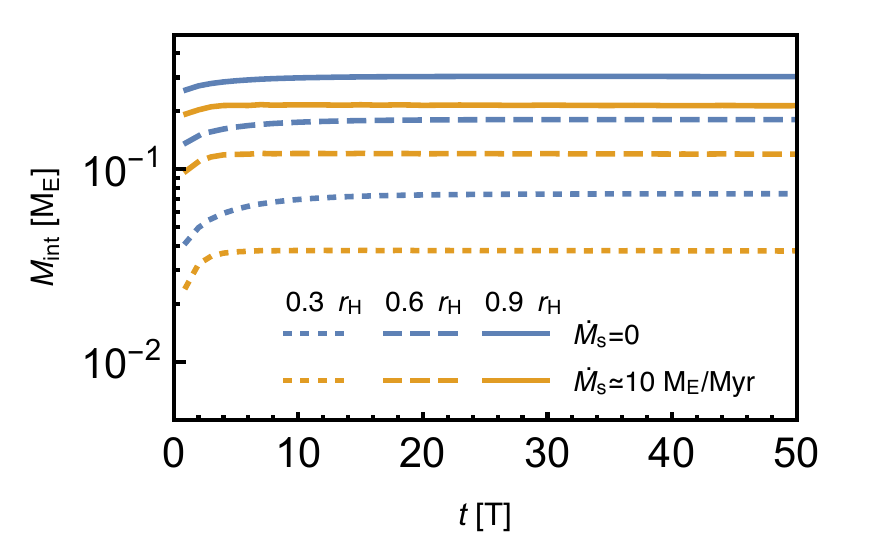}
  \caption{
	  Enclosed envelope mass as function of time. 
	  Orange lines corresponds to run \texttt{RADL27k0.01LONG}, blue lines
	  to \texttt{RADL0k0.01LONG}.
	  Short dashed, dashed and full curves correspond to the mass within a
	  radius of respectively $0.3,0.6,0.9$\,$r_{\rm H}$.
	  Convergence is reached within approximately $10$ orbits.
	  The inner envelope shells take the longest time to settle, especially
	  when pressure balance is driven by compressional heating.  
	  The potential is introduced during one orbital period
	  (Eq.\,\ref{eq:intropot}).
  }
  \label{fig:tmevol}
\end{figure}

\begin{figure}[t!]
  \centering
  \includegraphics[width=8.8cm]{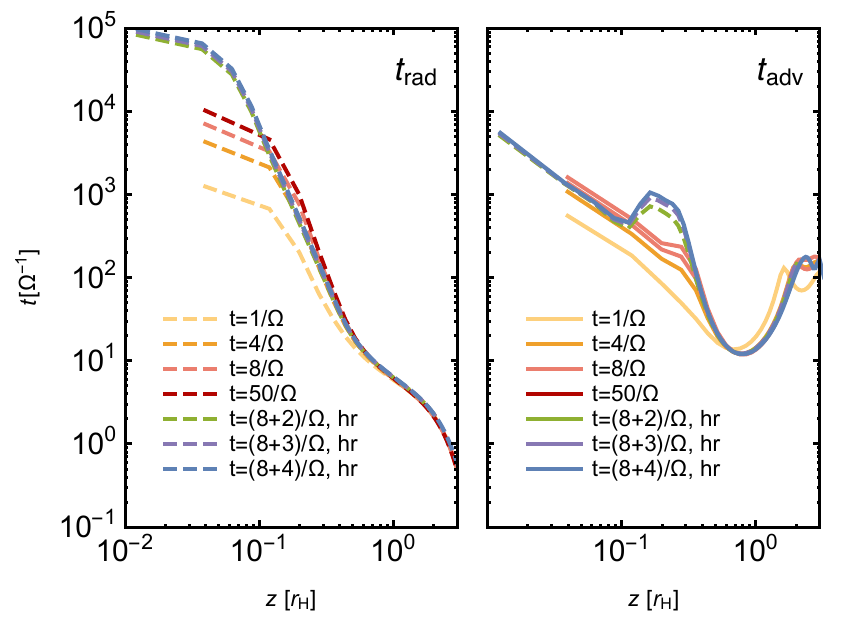}
  \caption{
	  Convergence of radiative and advective timescales. 
	  Radiative timescales in run \texttt{RADL27k0.01LONG} converge rapidly
	  in about $4$ orbits, outside the smoothing radius at $0.25\,r_{\rm
	  H}$. Little change occurs between $8$ to $50$ orbits.  
	  After $8$ medium resolution orbits, the restart at high resolution,
	  labeled with hr in the legend, does not alter $t_{\rm diff}$ outside
	  $0.1$\,$r_{\rm H}$.
	  Advective timescales also quickly settle. 
	  Note however that 
	  approximately $4$ orbits at high resolution 
	  are required to capture the 
	  increased vertical advection timescale around $0.2$\,$r_{\rm H}$
	  (run \texttt{RADL27k0.01}). 
  }
  \label{fig:tdiffconv}
\end{figure}

We have verified that we obtain steady state envelope structures.
Figure \ref{fig:tmevol} shows the evolution of the envelope mass with time. 
We find an envelope structure that remains stable on timescales of around $50$
orbits. We also verified this by inspecting the stability of the total torque. 
The long term integrations presented in Fig.\,\ref{fig:tmevol}, do not
reveal any accretion of gas.
Even longer simulations, exceeding $10^5$ years, would likely be necessary in
order to capture the secular cooling process of the envelope and the
resulting growth in envelope mass.
Inspection of Fig.\,\ref{fig:tdiffconv} shows that about $8$ orbits at medium
resolution are sufficient to capture the thermal structure and advection flow
inside the planetary envelope to sufficient precision.
However, we do note that high resolution simulations are required to capture
the flow structure around $r\approx0.2\,r_{\rm H}$. 
Therefore, about $4$ orbits at high resolution are required to capture the
effect of a reduced vertical flux.
Demonstrating full convergence in $t_{\rm adv}$ would demand running a high
resolution simulation for a time $t \gtrsim t_{\rm adv}$, which is currently
unfeasible. Nevertheless, since no qualitative changes are expected to occur
physically at later times, a shorter run time is sufficient in the context
of this work.

\section{Methods: Accretion luminosity}
\label{ap:L}

We use a simple prescription to model the accretion heat released at the
interior of the planetary envelope \citep{Benitez_2015}. 
All accretion heat $L_{\rm acc}$ is assumed to be deposited uniformly in
the volume prescribed by the inner grid cells around the planet core.
Simulations are mirrored across the midplane, so we only prescribe the $4$ grid
cells of the upper midplane that are centered around the midpoint of the
planetary potential.
For each of the $4$ cells, we set an energy deposit rate of 
\begin{align}
  l = \frac{L_{\rm acc}}{8 V_{\rm cell}}\,.
\end{align}
Here, $V_{\rm cell}$ is the grid cell volume, which is well described by a
constant, since our non-uniform grids are to a very good accuracy equal-sized
close to the position of the planet.

\section{Methods: Isothermal Equation of state}
\label{ap:iso}

\begin{figure}[t!]
  \centering
  \includegraphics[width=8.8cm]{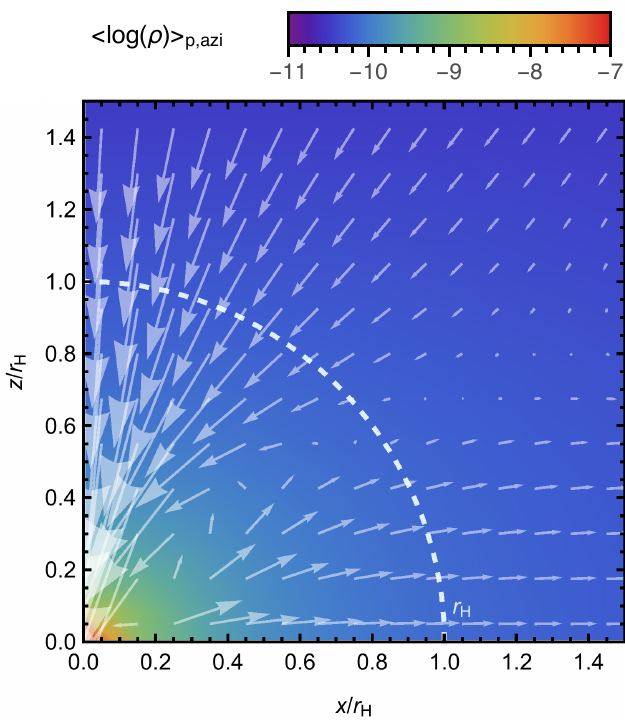}
  \caption{
  Azimuthally-averaged velocity field around a $5$\,M$_{\rm E}$-core, seen in the
  co-rotating frame, with the use of an isothermal equation of state
  (run \texttt{ISO}).  The background shows the azimuthally averaged gas density
  ($ \left< \log \rho \right>_{\rm p,azi} $ in g/cm$^3$).
  }
  \label{fig:stream_aziaver_iso}
\end{figure}

\begin{figure}[t!]
  \centering
  \includegraphics[width=8.8cm]{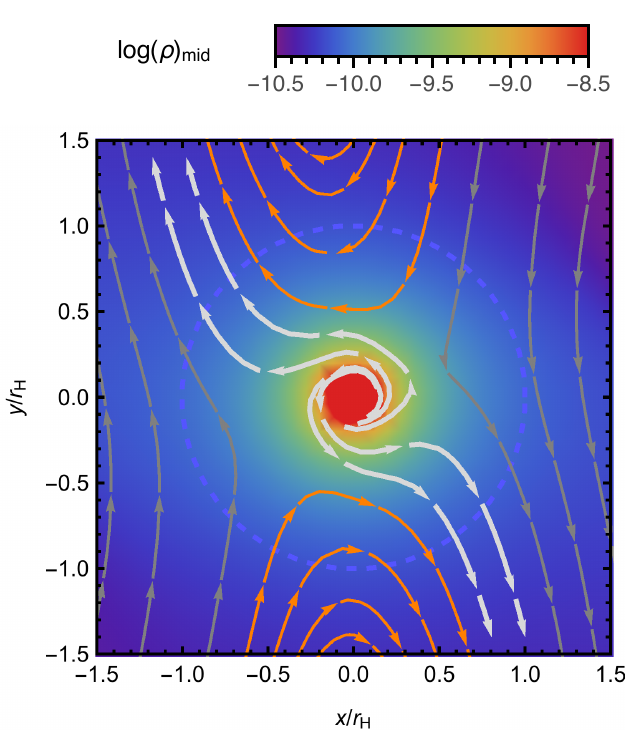}
  \caption{
  Velocity field around a $5$\,M$_{\rm E}$-core in the co-rotating frame (run
  \texttt{ISO}).
  Displayed are the streamlines in the midplane ($z=0$). One can identify the
  outer regions
  dominated by Keplerian shear, the horshoe orbits in front and behind the 
  planet, and two distinct arms transporting mass away from the planet. 
  The background shows the gas density in the midplane.
  }
  \label{fig:stream_midplane_iso}
\end{figure}

In this section, we compare \texttt{FARGOCA} using an isothermal equation of
state, with previous results obtained in the literature for the 3D flow in low
viscosity discs. In this way we test the code and more specifically the
implementation of the 3D non-uniform grid.

We first compare the morphology of the flow field in the disc midplane around
the planet. We have used a parameter set very close to the one reported by
\citet{Fung_2015} (simulation \texttt{ISO}).
Figure \ref{fig:stream_midplane_iso} and \ref{fig:stream_aziaver_iso} show the
motion of gas around the planet. 
We find the same morphology as reported in
\citet{Fung_2015}, for their 3D results in a low viscosity disc, and
\citet{Ormel_2015}.
Gas enter vertically the Hill sphere (Fig.\,\ref{fig:stream_aziaver_iso}) and
escapes along the midplane through two channels in the second and fourth
quadrant (white streamlines in Fig.\,\ref{fig:stream_midplane_iso}). This
represents the transient horseshoe flow reported in \citet{Fung_2015}.

\section{Hydrostatic model of a spherical atmosphere}
\label{ap:1D}

In this appendix, we briefly discuss the structure of a planetary envelope in
hydrostatic balance \citep{Mizuno_1980, Rafikov_2006, Piso_2014}.
In the outer shell, energy transport occurs through radiation. The
temperature gradient with respect to the pressure takes the form 
\begin{align}
  \nabla_{\rm rad} = \frac{d \ln T}{ d \ln P} 
  = \frac{3 \kappa(r) P(r) }{64 \pi G \sigma_{\rm SB} T(r)^4} 
  \frac{L(r)}{M_{\rm int}(r)} \,.
  \label{eq:nablarad}
\end{align}
We now will assume a constant opacity and luminosity through the envelope, 
$L(r)=L$. We consider low mass envelopes, therefore the potential is dominated
by the core, $M_{\rm int}(r) = M_{\rm c}$. 
Then the temperature remains nearly constant in the outer envelope, $T(r) =
T_{\rm out}$. Thus the pressure and density are exponentially increasing,
\begin{align}
  P = P_{\rm out} \exp \left[ r_{\rm B} \left(\frac{1}{r} - \frac{1}{r_{\rm
  out}} \right) \right] \,.
  \label{eq:expP}
\end{align}
Here $ P_{\rm out}$ is the pressure at an unperturbed point at $ r_{\rm out}$,
often taken to be the Bondi radius.

The temperature gradient cannot grow unbounded towards the interior. 
Steep temperature gradients become unstable to convection. 
Energy transport starts occurring through overturning motions of gas, that are
assumed to maintain a critical temperature gradient $\nabla_{\rm ad} =
\frac{\gamma-1}{\gamma}$.
Therefore the depth of the radiative zone is set by the point where 
$  \nabla_{\rm ad} = \nabla_{\rm rad}$. 
This allows us the rewrite Eq.\,\ref{eq:nablarad} to find the pressure at the
radiative convective boundary,
\begin{align}
  P_{\rm rcb} =  \frac{64 \pi G \sigma_{\rm SB}  \nabla_{\rm ad} T_{\rm out}^4
  M_{\rm c}}{3 \kappa L }\,.
  \label{eq:PrcbPmid}
\end{align}
Here the subscript ${\rm rcb}$ stands for values evaluated at the point where outward
radiation transport starts, at the radiative convective boundary. 
The depth of the radiative zone can be expressed, using Eq.\,\ref{eq:expP}, as
\begin{align}
  \frac{r_{\rm rcb}}{r_{\rm out}} 
  &\approx \left[ \ln \left( \frac{P_{\rm rcb}}{P_{\rm B}}
  \right) + \frac{r_{\rm B}}{r_{\rm out}} \right]^{-1}  \\
  &= \left[ \ln \left(
    \frac{64 \pi G \sigma_{\rm SB}  \nabla_{\rm ad} T_{\rm out}^4 M_{\rm core}}{3
    \kappa L P_{\rm out} }
    \right)  + \frac{r_{\rm B}}{r_{\rm out}} \right]^{-1} \,.
\end{align}
This expression represents a powerful relationship between the depth of the
radiative zone, the opacity and the luminosity of the planet.
Note also that the depth of the radiative zone uniquely determines the
energy stored in the envelope. Hot, nearly fully convective envelopes, have
$r_{\rm rcb}$ close to the outer boundary edge of the atmosphere. 
Cooling down the envelope goes paired with the radiative zone entering
progressively deeper in the envelope.

\section{Pebble accretion}
\label{app:pa}

\begin{figure}[t!]
  \centering
  \includegraphics[width=8.8cm]{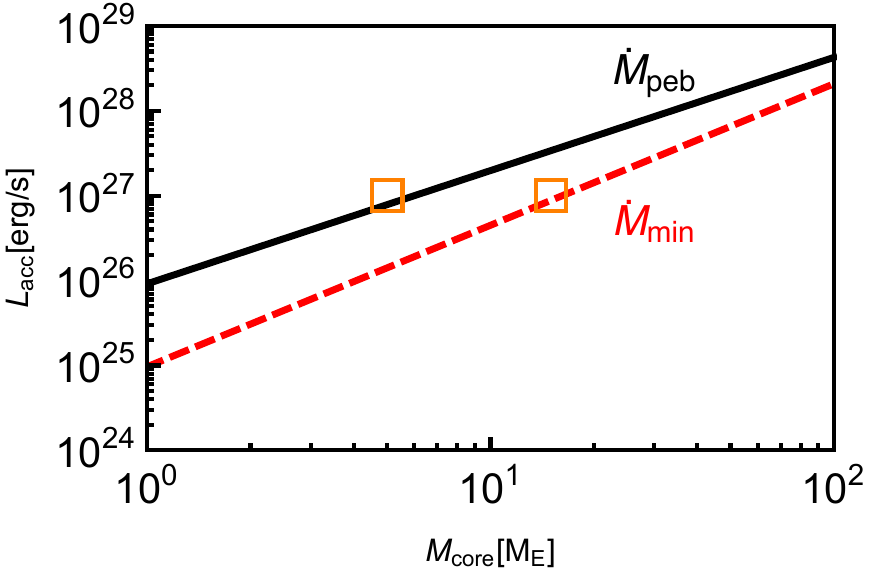}
  \caption{
	  Luminosity from the accretion of solids.
	  The black curve corresponds to the luminosity by pebble accretion,
	  which can be considered the maximal possible accretion rate.
	  The red dashed line gives a lower bound on the accretion luminosity
	  from the requirement order to have sufficiently high accretion rates
	  to finish core growth within the lifetime of the protoplanetary disc.
	  The orange squares represent the accretion luminosities explored in
	  this work. 
  }
  \label{fig:Lacc}
\end{figure}

In Fig.\,\ref{fig:Lacc}, we present the accretion luminosity for a core growing
by the accretion of pebbles, in a nominal pebble disc model
\citep{Lambrechts_2014b}.
The efficient accretion of 
drag-sensitive particles in the cm size range allows cores to grow to
completion before gas disc dissipation.
This growth rate can be considered to give an upper limit on the accretion
luminosity, because pebbles be can accreted from the full Hill sphere and the
whole mass reservoir of inwards drifting pebbles is available.
Conversely, accretion rates cannot have been much lower in order to meet the
constraint that cores form before disc dissipation. 
The red curve in Fig.\,\ref{fig:Lacc} is constructed by assuming the upper bound
on the growth timescale of $M_{\rm c}/\dot M_{\rm min} = 10$\,Myr, which
would correspond to a long lived protoplanetary disc.

\bibliographystyle{aa}        
\bibliography{references}     

\begin{thebibliography}{56}
\expandafter\ifx\csname natexlab\endcsname\relax\def\natexlab#1{#1}\fi

\bibitem[{{Ayliffe} \& {Bate}(2012)}]{Ayliffe_2012}
{Ayliffe}, B.~A. \& {Bate}, M.~R. 2012, \mnras, 427, 2597

\bibitem[{{Bell} \& {Lin}(1994)}]{Bell_1994}
{Bell}, K.~R. \& {Lin}, D.~N.~C. 1994, \apj, 427, 987

\bibitem[{{Ben{\'{\i}}tez-Llambay} {et~al.}(2015){Ben{\'{\i}}tez-Llambay},
  {Masset}, {Koenigsberger}, \& {Szul{\'a}gyi}}]{Benitez_2015}
{Ben{\'{\i}}tez-Llambay}, P., {Masset}, F., {Koenigsberger}, G., \&
  {Szul{\'a}gyi}, J. 2015, \nat, 520, 63

\bibitem[{{Bitsch} {et~al.}(2013){Bitsch}, {Crida}, {Morbidelli}, {Kley}, \&
  {Dobbs-Dixon}}]{Bitsch_2013}
{Bitsch}, B., {Crida}, A., {Morbidelli}, A., {Kley}, W., \& {Dobbs-Dixon}, I.
  2013, \aap, 549, A124

\bibitem[{{Bitsch} {et~al.}(2015){Bitsch}, {Lambrechts}, \&
  {Johansen}}]{Bitsch_2015}
{Bitsch}, B., {Lambrechts}, M., \& {Johansen}, A. 2015, \aap, 582, A112

\bibitem[{{Brauer} {et~al.}(2008){Brauer}, {Dullemond}, \&
  {Henning}}]{Brauer_2008}
{Brauer}, F., {Dullemond}, C.~P., \& {Henning}, T. 2008, \aap, 480, 859

\bibitem[{{Buchhave} {et~al.}(2012){Buchhave}, {Latham}, {Johansen},
  {Bizzarro}, {Torres}, {Rowe}, {Batalha}, {Borucki}, {Brugamyer}, {Caldwell},
  {Bryson}, {Ciardi}, {Cochran}, {Endl}, {Esquerdo}, {Ford}, {Geary},
  {Gilliland}, {Hansen}, {Isaacson}, {Laird}, {Lucas}, {Marcy}, {Morse},
  {Robertson}, {Shporer}, {Stefanik}, {Still}, \& {Quinn}}]{Buchhave_2012}
{Buchhave}, L.~A., {Latham}, D.~W., {Johansen}, A., {et~al.} 2012, \nat, 486,
  375

\bibitem[{{Commer{\c c}on} {et~al.}(2011){Commer{\c c}on}, {Teyssier}, {Audit},
  {Hennebelle}, \& {Chabrier}}]{Commercon_2011}
{Commer{\c c}on}, B., {Teyssier}, R., {Audit}, E., {Hennebelle}, P., \&
  {Chabrier}, G. 2011, \aap, 529, A35

\bibitem[{{D'Angelo} \& {Bodenheimer}(2013)}]{Angelo_2013}
{D'Angelo}, G. \& {Bodenheimer}, P. 2013, \apj, 778, 77

\bibitem[{{D'Angelo} \& {Lubow}(2008)}]{Angelo_2008}
{D'Angelo}, G. \& {Lubow}, S.~H. 2008, \apj, 685, 560

\bibitem[{{de Val-Borro} {et~al.}(2006){de Val-Borro}, {Edgar}, {Artymowicz},
  {Ciecielag}, {Cresswell}, {D'Angelo}, {Delgado-Donate}, {Dirksen}, {Fromang},
  {Gawryszczak}, {Klahr}, {Kley}, {Lyra}, {Masset}, {Mellema}, {Nelson},
  {Paardekooper}, {Peplinski}, {Pierens}, {Plewa}, {Rice}, {Sch{\"a}fer}, \&
  {Speith}}]{deVal_2006}
{de Val-Borro}, M., {Edgar}, R.~G., {Artymowicz}, P., {et~al.} 2006, \mnras,
  370, 529

\bibitem[{{Fung} {et~al.}(2015){Fung}, {Artymowicz}, \& {Wu}}]{Fung_2015}
{Fung}, J., {Artymowicz}, P., \& {Wu}, Y. 2015, \apj, 811, 101

\bibitem[{{Gressel} {et~al.}(2013){Gressel}, {Nelson}, {Turner}, \&
  {Ziegler}}]{Gressel_2013}
{Gressel}, O., {Nelson}, R.~P., {Turner}, N.~J., \& {Ziegler}, U. 2013, \apj,
  779, 59

\bibitem[{{Hadden} \& {Lithwick}(2016)}]{Hadden_2016}
{Hadden}, S. \& {Lithwick}, Y. 2016, ArXiv e-prints

\bibitem[{{Hori} \& {Ikoma}(2011)}]{Hori_2011}
{Hori}, Y. \& {Ikoma}, M. 2011, \mnras, 416, 1419

\bibitem[{{Hubickyj} {et~al.}(2005){Hubickyj}, {Bodenheimer}, \&
  {Lissauer}}]{Hubickyj_2005}
{Hubickyj}, O., {Bodenheimer}, P., \& {Lissauer}, J.~J. 2005, \icarus, 179, 415

\bibitem[{{Ikoma} {et~al.}(2000){Ikoma}, {Nakazawa}, \& {Emori}}]{Ikoma_2000}
{Ikoma}, M., {Nakazawa}, K., \& {Emori}, H. 2000, \apj, 537, 1013

\bibitem[{{Klahr} \& {Kley}(2006)}]{Klahr_2006}
{Klahr}, H. \& {Kley}, W. 2006, \aap, 445, 747

\bibitem[{{Kley}(1989)}]{Kley_1989}
{Kley}, W. 1989, \aap, 208, 98

\bibitem[{{Lambrechts} \& {Johansen}(2012)}]{Lambrechts_2012}
{Lambrechts}, M. \& {Johansen}, A. 2012, \aap, 544, A32

\bibitem[{{Lambrechts} \& {Johansen}(2014)}]{Lambrechts_2014b}
{Lambrechts}, M. \& {Johansen}, A. 2014, \aap, 572, A107

\bibitem[{{Lambrechts} {et~al.}(2016){Lambrechts}, {Johansen}, {Capelo},
  {Blum}, \& {Bodenschatz}}]{Lambrechts_2016}
{Lambrechts}, M., {Johansen}, A., {Capelo}, H.~L., {Blum}, J., \&
  {Bodenschatz}, E. 2016, \aap, 591, A133

\bibitem[{{Lambrechts} {et~al.}(2014){Lambrechts}, {Johansen}, \&
  {Morbidelli}}]{Lambrechts_2014a}
{Lambrechts}, M., {Johansen}, A., \& {Morbidelli}, A. 2014, \aap, 572, A35

\bibitem[{{Lee} {et~al.}(2014){Lee}, {Chiang}, \& {Ormel}}]{Lee_2014}
{Lee}, E.~J., {Chiang}, E., \& {Ormel}, C.~W. 2014, \apj, 797, 95

\bibitem[{{Lega} {et~al.}(2014){Lega}, {Crida}, {Bitsch}, \&
  {Morbidelli}}]{Lega_2014}
{Lega}, E., {Crida}, A., {Bitsch}, B., \& {Morbidelli}, A. 2014, \mnras, 440,
  683

\bibitem[{{Lega} {et~al.}(2015){Lega}, {Morbidelli}, {Bitsch}, {Crida}, \&
  {Szul{\'a}gyi}}]{Lega_2015}
{Lega}, E., {Morbidelli}, A., {Bitsch}, B., {Crida}, A., \& {Szul{\'a}gyi}, J.
  2015, \mnras, 452, 1717

\bibitem[{{Levermore} \& {Pomraning}(1981)}]{Levermore_1981}
{Levermore}, C.~D. \& {Pomraning}, G.~C. 1981, \apj, 248, 321

\bibitem[{{Levison} {et~al.}(2010){Levison}, {Thommes}, \&
  {Duncan}}]{Levison_2010}
{Levison}, H.~F., {Thommes}, E., \& {Duncan}, M.~J. 2010, \aj, 139, 1297

\bibitem[{{Lissauer} {et~al.}(2009){Lissauer}, {Hubickyj}, {D'Angelo}, \&
  {Bodenheimer}}]{Lissauer_2009}
{Lissauer}, J.~J., {Hubickyj}, O., {D'Angelo}, G., \& {Bodenheimer}, P. 2009,
  \icarus, 199, 338

\bibitem[{{Machida} {et~al.}(2008){Machida}, {Kokubo}, {Inutsuka}, \&
  {Matsumoto}}]{Machida_2008}
{Machida}, M.~N., {Kokubo}, E., {Inutsuka}, S.-i., \& {Matsumoto}, T. 2008,
  \apj, 685, 1220

\bibitem[{{Machida} {et~al.}(2010){Machida}, {Kokubo}, {Inutsuka}, \&
  {Matsumoto}}]{Machida_2010}
{Machida}, M.~N., {Kokubo}, E., {Inutsuka}, S.-I., \& {Matsumoto}, T. 2010,
  \mnras, 405, 1227

\bibitem[{{Masset}(2000)}]{Masset_2000}
{Masset}, F. 2000, \aaps, 141, 165

\bibitem[{{Mihalas} \& {Mihalas}(1984)}]{Mihalas_1984}
{Mihalas}, D. \& {Mihalas}, B.~W. 1984, {Foundations of radiation
  hydrodynamics}

\bibitem[{{Mizuno}(1980)}]{Mizuno_1980}
{Mizuno}, H. 1980, Progress of Theoretical Physics, 64, 544

\bibitem[{{Morbidelli} {et~al.}(2014){Morbidelli}, {Szul{\'a}gyi}, {Crida},
  {Lega}, {Bitsch}, {Tanigawa}, \& {Kanagawa}}]{Morbidelli_2014}
{Morbidelli}, A., {Szul{\'a}gyi}, J., {Crida}, A., {et~al.} 2014, \icarus, 232,
  266

\bibitem[{{Mordasini}(2014)}]{Mordasini_2014}
{Mordasini}, C. 2014, \aap, 572, A118

\bibitem[{{Movshovitz} {et~al.}(2010){Movshovitz}, {Bodenheimer}, {Podolak}, \&
  {Lissauer}}]{Movshovitz_2010}
{Movshovitz}, N., {Bodenheimer}, P., {Podolak}, M., \& {Lissauer}, J.~J. 2010,
  \icarus, 209, 616

\bibitem[{{Ormel}(2014)}]{Ormel_2014}
{Ormel}, C.~W. 2014, \apjl, 789, L18

\bibitem[{{Ormel} \& {Klahr}(2010)}]{Ormel_2010}
{Ormel}, C.~W. \& {Klahr}, H.~H. 2010, \aap, 520, A43

\bibitem[{{Ormel} {et~al.}(2015){Ormel}, {Shi}, \& {Kuiper}}]{Ormel_2015}
{Ormel}, C.~W., {Shi}, J.-M., \& {Kuiper}, R. 2015, \mnras, 447, 3512

\bibitem[{{Piso} \& {Youdin}(2014)}]{Piso_2014}
{Piso}, A.-M.~A. \& {Youdin}, A.~N. 2014, \apj, 786, 21

\bibitem[{{Pollack} {et~al.}(1996){Pollack}, {Hubickyj}, {Bodenheimer},
  {Lissauer}, {Podolak}, \& {Greenzweig}}]{Pollack_1996}
{Pollack}, J.~B., {Hubickyj}, O., {Bodenheimer}, P., {et~al.} 1996, \icarus,
  124, 62

\bibitem[{{Rafikov}(2004)}]{Rafikov_2004}
{Rafikov}, R.~R. 2004, \aj, 128, 1348

\bibitem[{{Rafikov}(2006)}]{Rafikov_2006}
{Rafikov}, R.~R. 2006, \apj, 648, 666

\bibitem[{{Ribas} {et~al.}(2015){Ribas}, {Bouy}, \&
  {Mer{\'{\i}}n}}]{Ribas_2015}
{Ribas}, {\'A}., {Bouy}, H., \& {Mer{\'{\i}}n}, B. 2015, \aap, 576, A52

\bibitem[{{Sato} {et~al.}(2016){Sato}, {Okuzumi}, \& {Ida}}]{Sato_2016}
{Sato}, T., {Okuzumi}, S., \& {Ida}, S. 2016, \aap, 589, A15

\bibitem[{{Semenov} {et~al.}(2003){Semenov}, {Henning}, {Helling}, {Ilgner}, \&
  {Sedlmayr}}]{Semenov_2003}
{Semenov}, D., {Henning}, T., {Helling}, C., {Ilgner}, M., \& {Sedlmayr}, E.
  2003, \aap, 410, 611

\bibitem[{{Stevenson}(1982)}]{Stevenson_1982}
{Stevenson}, D.~J. 1982, \planss, 30, 755

\bibitem[{{Stone} \& {Norman}(1992)}]{Stone_1992}
{Stone}, J.~M. \& {Norman}, M.~L. 1992, \apjs, 80, 753

\bibitem[{{Szul{\'a}gyi} {et~al.}(2014){Szul{\'a}gyi}, {Morbidelli}, {Crida},
  \& {Masset}}]{Szul_2014}
{Szul{\'a}gyi}, J., {Morbidelli}, A., {Crida}, A., \& {Masset}, F. 2014, \apj,
  782, 65

\bibitem[{{Tanaka} \& {Ida}(1999)}]{Tanaka_1999}
{Tanaka}, H. \& {Ida}, S. 1999, \icarus, 139, 350

\bibitem[{{Tanigawa} {et~al.}(2012){Tanigawa}, {Ohtsuki}, \&
  {Machida}}]{Tanigawa_2012}
{Tanigawa}, T., {Ohtsuki}, K., \& {Machida}, M.~N. 2012, \apj, 747, 47

\bibitem[{{Thorngren} {et~al.}(2016){Thorngren}, {Fortney}, {Murray-Clay}, \&
  {Lopez}}]{Thorngren_2016}
{Thorngren}, D.~P., {Fortney}, J.~J., {Murray-Clay}, R.~A., \& {Lopez}, E.~D.
  2016, \apj, 831, 64

\bibitem[{{Venturini} {et~al.}(2016){Venturini}, {Alibert}, \&
  {Benz}}]{Venturini_2016}
{Venturini}, J., {Alibert}, Y., \& {Benz}, W. 2016, \aap, 596, A90

\bibitem[{{Venturini} {et~al.}(2015){Venturini}, {Alibert}, {Benz}, \&
  {Ikoma}}]{Venturini_2015}
{Venturini}, J., {Alibert}, Y., {Benz}, W., \& {Ikoma}, M. 2015, \aap, 576,
  A114

\bibitem[{{Winn} \& {Fabrycky}(2015)}]{Winn_2015}
{Winn}, J.~N. \& {Fabrycky}, D.~C. 2015, \araa, 53, 409

\end{thebibliography}

\end{document}